\documentclass[journal,comsoc]{IEEEtran} %

\usepackage[T1]{fontenc}%
\usepackage[noadjust]{cite}

\ifCLASSINFOpdf
   \usepackage[pdftex]{graphicx}
  \graphicspath{{figures/drawings/}{figures/plots/pdf/}}
\else
\fi
\usepackage{amsmath}
\usepackage[cmintegrals]{newtxmath}
\ifCLASSOPTIONcompsoc
  \usepackage[caption=false,font=normalsize,labelfont=sf,textfont=sf]{subfig}
\else
  \usepackage[caption=false,font=footnotesize]{subfig}
\fi
\begin{document}
\title{Open Set Wireless Transmitter Authorization: Deep Learning Approaches and Dataset Considerations}

\author{Samer~Hanna,~\IEEEmembership{Student~Member,~IEEE,}
	Samurdhi~Karunaratne,~\IEEEmembership{Student~Member,~IEEE,}
	and Danijela~Cabric,~\IEEEmembership{Senior~Member,~IEEE}%
	\thanks{Authors are with the Electrical and Computer Engineering Department, University of California, Los Angeles, CA 90095, USA.
	 e-mails: 	\mbox{samerhanna@ucla.edu}, samurdhi@ucla.edu, danijela@ee.ucla.edu}
	\thanks{This work was supported in part by the CONIX Research Center, one of six centers in JUMP, a Semiconductor Research Corporation (SRC) program sponsored by DARPA.}
	}
\maketitle

\begin{abstract}
Due to  imperfections in transmitters' hardware, wireless signals can be used to  verify their identity in an authorization system.  While deep learning was proposed for transmitter identification, the majority of the work has focused on  classification among a closed set of transmitters.  Malicious transmitters outside this closed set will be misclassified, jeopardizing the authorization system.
In this paper, we consider the problem of recognizing authorized transmitters and rejecting new  transmitters. To address this problem, we adapt the most prominent approaches from the open set recognition and anomaly detection literature to the problem. We study how these approaches scale with the required number of authorized transmitters. We propose using a known set of unauthorized transmitters to assist the training and study its impact. The evaluation procedure takes into consideration that some transmitters might be more similar than others and nuances these effects. The robustness of the RF authorization  with respect to temporal changes in fingerprints is also considered in the evaluation.  When using 10 authorized  and 50 known unauthorized WiFi transmitters from a publicly accessible testbed, we were able to achieve an outlier detection accuracy of 98\% on the same day test set and 80\% on the different day test set.
\end{abstract}


\IEEEpeerreviewmaketitle

\section{Introduction}

With the growth in the number of wireless connected devices, securing  them has become more challenging. Unlike wired communications, a wireless network is accessible by any device with sufficient transmit power. This makes  \textit{authentication}, the process of verifying the identity of devices, challenging. After authentication, devices are granted access, the process known a \textit{authorization}. While cryptographic methods are used for authentication, many devices like Internet-of-Things (IoT) devices don't possess the energy nor computational power to run them, leading to  many authentication based attacks~\cite{neshenko_demystifying_2019}.

Physical Layer Authentication (PLA) leverages the dynamics of physical layer attributes to address these challenges and to enhance wireless security~\cite{wang_physical-layer_2016}. While active PLA typically require changes in transmitters, passive PLA is performed  on the receiver side, making it more practical. Passive PLA uses RF fingerprints; combining  channel state information (CSI) and  transmitter hardware fingerprints to authenticate devices. Transmitter fingerprints result from the imperfections in their RF chain components like ADCs, power amplifiers, etc. The interaction between these non-idealities makes signals from identical transmitters exhibit unique characteristics typically modeled as carrier frequency offset, IQ imbalance, among others~\cite{wang_wireless_2016}.  %

While there has been many approaches for using RF fingerprints based on  handcrafted features  \cite{yu_blind_2016,brik_wireless_2008,danev_transient-based_2009,vo-huu_fingerprinting_2016,ren_practical_2018,peng_design_2019,chatterjee_rf-puf_2018,zhou_design_2019,xiao_using_2008,senigagliesi_statistical_2019,zhang_physical_2019}, it was shown to be highly dependent on the quality of the receiver hardware \cite{rehman_analysis_2012} and requires manual feature engineering which are protocol dependent. For these reasons, recently, there has been wide interest is using deep learning approaches to address this problem \cite{riyaz_deep_2018,yu_robust_2019,gopalakrishnan_robust_2019,baldini_comparison_2019,agadakos_deep_2019,wu_deep_2018,merchant_deep_2018-1,hanna_icnc_2019,youssef_machine_2017,gritsenko_finding_2019,hanna_spawc_2020}. Deep learning has the ability to learn a richer set of features from raw IQ samples leading to improved performance over manually selected features, as has been demonstrated in~\cite{riyaz_deep_2018}.%

While previous deep learning work in this area has addressed many of the challenges of RF fingerprinting, this body of work has posed the problem as a \textit{closed set} classification which assumes a  known set of transmitters, except for \cite{gritsenko_finding_2019} and our prior work \cite{hanna_spawc_2020}. No matter how large the set is, if any new unseen transmitter gets within communication range, its signal will get misclassified leading to security vulnerabilities. This calls for  \textit{open set} approaches which are capable of rejecting signals from unseen transmitters. While \cite{gritsenko_finding_2019} proposed their own approach to address this problem, rejecting samples  from a new distribution is not a novel problem for the machine learning community. A plethora of approaches  have already been proposed for similar problems in computer vision, natural language processing, intrusion detection, etc. Two problems are most relevant;  openset classification~\cite{openset_survey_2019}: classifying among known classes  and rejecting unseen classes, and \cite{anomaly_detection_survey_2019}: identifying abnormal samples. Instead of reinventing the wheel, we aim to  adapt the most prominent approaches for these problems and evaluate their performance. 
Unlike other domains, transmitter authorization has its own challenges and requirements: (1)  RF fingerprints arise from random channel and hardware variations, hence, generalizable conclusions can not be derived from single point evaluations (2) the number of authorized transmitters is a system requirement that  can vary significantly (3) the ease of collecting data (compared to image classification, for instance) raises questions about how to construct a training dataset, and (4)  the robustness of the approach against time varying fingerprints needs to be evaluated.

   In our previous work~\cite{hanna_spawc_2020}, which we extend in this work, we started investigating  a few approaches from the existing openset recognition  literature. In this work, our contributions can be summarized as follows:
\begin{itemize}
	\item  We formulate the problem of rejecting signals from unseen transmitters as both an openset recognition problem and an anomaly detection problem. Then we adapt and evaluate several well-established approaches to our problem. 
	\item We discuss several considerations for network architecture design for transmitter authorization. We show that making minor changes to the neural network architecture and data labeling strategy yields a conceptually different approach with different performance. We show that classification within a closed set is not always an indicator for performance in an open set. 
	\item We compare the performance of different approaches with respect to  the number of authorized transmitters required by the system. We propose using a set of known unauthorized transmitters and show its benefits.
	\item  We show that the results obtained are dependent on the choice of transmitters and time of evaluation. Then we address this by showing results in terms of statistics of multiple transmitter choices using data captures on same day as  training  and different day. 
\end{itemize}

\section{Related Work}
 Physical Layer Authentication (PLA) can be classified as active or passive. Active PLA typically overlays a tag over the message used for authentication, thus requiring changes to the physical layer of the transmitters~\cite{xie_blind_2018,gu_physical_2020}. Passive PLA on the other hand uses the channel state information and the RF fingerprint due to hardware imperfections to identify transmitters~\cite{wang_wireless_2016}, requiring no change to transmitter signals, and hence is easier to apply. Approaches for passive PLA either use a set of handcrafted features  or use deep learning directly on IQ signals.

\subsubsection{Handcrafted Feature PLA}
Feature based PLA has either considered transmitter fingerprints or channel state information to distinguish between transmitters.
A variety of features were considered as transmitter fingerprints  in the literature~\cite{xu_device_2016-1}. These features  include transient ones like the patterns at the start of packets~\cite{danev_transient-based_2009}, and steady-state ones like carrier frequency offset~\cite{yu_blind_2016}, IQ imbalance, sampling frequency offset  or a combination of these features \cite{brik_wireless_2008,vo-huu_fingerprinting_2016,ren_practical_2018,peng_design_2019,chatterjee_rf-puf_2018,zhou_design_2019}.

Other works have based their features on channel state information (CSI). This kind of approach has been considered for SISO  ~\cite{xiao_using_2008,senigagliesi_statistical_2019} and  massive MIMO \cite{zhang_physical_2019} communications.  Combining  CSI with transmitter fingerprints has also been proposed~\cite{fang_learning-aided_2018}.
Gaussian Mixture Models ~\cite{nguyen_device_2011,xiao_using_2008,gulati_gmm_2013} and statistical hypothesis testing~\cite{weinand_application_2017}  were proposed for transmitter authorization using a set of manually designed features.

\subsubsection{Deep Learning Based PLA}
In contrast to handcrafted features, deep learning approaches are able to extract better features from the high dimensional signals, thus leading to higher accuracy compared to feature-based approaches \cite{riyaz_deep_2018}, gaining widespread interest recently~\cite{riyaz_deep_2018,yu_robust_2019,gopalakrishnan_robust_2019,baldini_comparison_2019,agadakos_deep_2019,wu_deep_2018,merchant_deep_2018-1,hanna_icnc_2019,youssef_machine_2017}.   Some of these works fall under the category of active PLA, requiring changes in the signal, while others are passive.

In active approaches,  modifications are intentionally added to the signal to improve classification. A protocol  inserting IQ imbalance and DC offset impairments to improve the RF fingerprints was proposed in \cite{sankhe_oracle_2018,sankhe_no_2019}. FIR filtering was also considered in \cite{restuccia_deepradioid_2019} to optimize RF fingerprints. %
The majority of the work falls under the passive category and has focused on the data representation, the network type, or a specific transmitter characteristic.
\paragraph{Data representation}
The work in \cite{baldini_comparison_2019} has compared different data representations like wavelet transform and Short Time Fourier transform while in \cite{baldini_physical_2019}, the authors considered recurrence plots.   Applying the Hilbert-Huang transform to the signal and deep residual networks were proposed in \cite{pan_specific_2019}. Higher order statistics like bispectrum were proposed in \cite{ding_specific_2018-1}.

\paragraph{Network Architecture}
In \cite{youssef_machine_2017}, authors compared different types of neural networks and machine learning techniques.   CNNs and RNNs to classify IoT devices over a wide range of SNR in~\cite{jafari_iot_2018}. In ~\cite{merchant_deep_2018-1}, multiple CNN architectures were tested on indoor and outdoor data with a focus on cognitive radio applications.  Complex neural networks were proposed in  \cite{agadakos_deep_2019} using convolutional and recurrent architectures. 
 The effect of a dynamic channel on deep learning RF fingerprinting along with the type of data was the focus of the work in \cite{morin_transmitter_2019}. 
 In  \cite{yu_robust_2019}, the authors have considered a multisampling neural network using LOS and NLOS datasets.  Denoising autoencoders were also proposed for the same problem \cite{yu_radio_2019}.  
  
  \paragraph{Transmitter Characteristic} Some works used deep learning while focusing on a specific type of impairment. The effect of power amplifier nonlinearity and signal type on classification performance was studied in \cite{hanna_icnc_2019}. In \cite{wong_emitter_2018}, CNNs were used to learn IQ imbalance as a modulation-independent way of transmitter identification.

The main limitation of this body of work is that it focuses on classification among a closed set of known transmitters. To the best of the authors' knowledge, two works have considered the problem of using deep learning for transmitter authorization that generalizes to unseen transmitters. The first work has proposed a novel approach for outlier detection that works on a per-packet basis~\cite{gritsenko_finding_2019}. The classifying neural network is applied to slices of the packet and statistics of the slices predictions are compared to a threshold. This approach is discussed further later in this work in Section~\ref{subsubsec:openmax}.
In their work, several datasets using WiFi and ADS-B were considered using 50, 250, and 500 devices. The data is said to have been captured "in the wild" with no further details provided. %
We only mention their results most similar to our work; using 50 authorized WiFi devices, they were able to detect new devices with an accuracy of 73\% at the cost of  a drop in classification accuracy from 63\%  to 43\%.

The second work that has considered this problem was our previous work~\cite{hanna_spawc_2020}, which is extended in this work. In~\cite{hanna_spawc_2020}, three approaches based on open set recognition were contrasted using a dataset consisting of WiFi preambles captured in a publicly accessible wireless testbed~\cite{orbit_2005}.   We have considered the effects of the number of authorized transmitters and demonstrated the benefit of using known outliers in training. Using 40 authorized transmitters, we were able to identify new devices from authorized devices with an accuracy of 84\%. The improved results obtained in our work can be partially attributed to the way the training and testing samples were generated, as discussed in Section~\ref{sec:dataset}. 

Compared to~\cite{hanna_spawc_2020}, in this work,  we consider more approaches that help to improve transmitter authorization with fewer transmitters. More analysis and insight is provided for the results. We further explore considerations for neural network architecture design with regard to outlier detection and how it differs from classification. The dataset used is expanded to include more transmitters, captured over a period of five days, increasing the confidence of our results and exploring temporal generalization.

\section{Classification, Openset Recognition, and Anomaly Detection}
\label{sec:class_openset_anomaly}
\begin{figure*}[t]
	\centering
	\subfloat[Classification \label{fig:ex_class}]{\includegraphics[scale = 0.8]{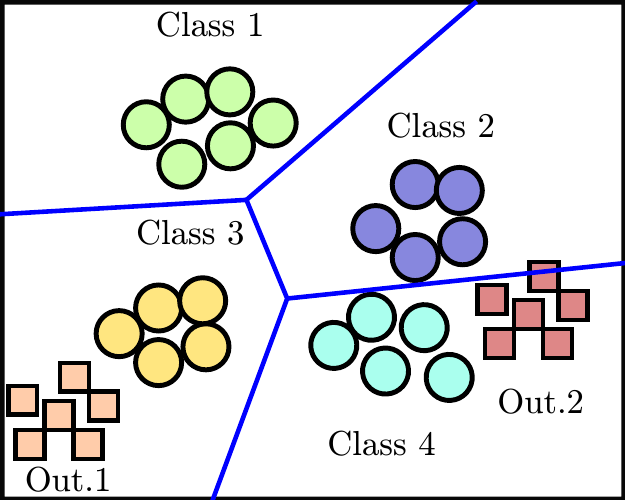}}  \hspace{5mm}
	\subfloat[Anomaly Detection \label{fig:ex_anomaly}]{\includegraphics[scale = 0.8]{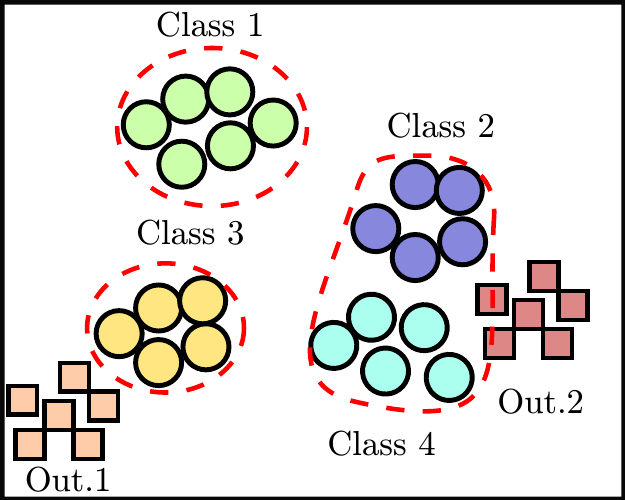}} \hspace{5mm}
	\subfloat[Openset Recognition \label{fig:ex_openset}]{\includegraphics[scale = 0.8]{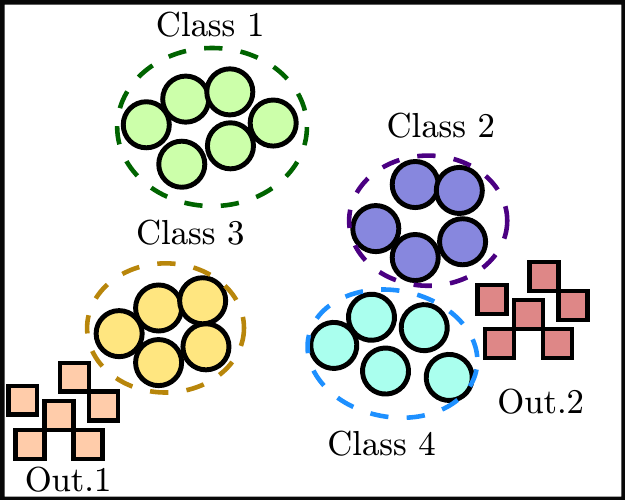}}  
	\caption{Known classes are depicted as circles and outlier classes as squares. Classifiers would mistakenly label oultiers. Anomaly detector rejects outliers but cannot distinguish among the known classes. Openset classification classifies among known classes and rejects outliers.}
	\label{fig:openset_anomaly_class}
\end{figure*}
 In this section, we highlight the difference between  classification,  openset recognition, and anomaly detection. A closed set classifier  determines boundaries that separate a pre-defined finite set of classes, shown as colored circles in Fig.~\ref{fig:ex_class}. As such, for samples from new classes (shown as squares in Fig.~\ref{fig:ex_class}) the classifier will predict the nearest class. This poses a grave security risk for a wireless authentication system. Since, it is impossible to train a classifier on all the transmitters in the world,  an approach that generalizes to signals from new unseen transmitters is needed.
 
We consider two formulations to address this limitation. The first one is posing the problem as an anomaly detection~\cite{anomaly_detection_survey_2019}. Anomaly detection aims to identify instances which are different from the normal.  %
 Hence, it finds boundaries around the seen classes (considered as normal cases), as shown in Fig.~\ref{fig:ex_anomaly}. The limitation, however, is that it treats all authorized transmitters as a single class.  An  overestimation of the determined boundaries could lead to errors in outlier detection as illustrated for outlier 2 in Fig.~\ref{fig:ex_anomaly}.
Open set classification~\cite{openset_survey_2019} takes an approach similar to anomaly detection while additionally classifying among the known samples. Hence, it isolates each class on its own as shown in Fig.~\ref{fig:ex_openset}. 

Classifying among transmitters using only received signals in a robust manner is a challenging problem on its own.  Extending it to open set poses even more challenges.
Unlike  approaches using handcrafted features \cite{gulati_gmm_2013,xu_device_2016-1,xiao_using_2008,nguyen_device_2011}, which use well separated features like CSI, in order to leverage the power of deep learning, we use raw IQ samples. The challenge for the deep neural networks is to learn features which separate the known classes from the unknown classes, for which no training samples are available. %

\providecommand{\mA}{\mathcal{A} }
\providecommand{\mK}{\mathcal{K} }
\providecommand{\mO}{\mathcal{O} }

\providecommand{\mAc}{\mathcal{|A|} }
\providecommand{\mKc}{\mathcal{|K|} }
\providecommand{\mOc}{\mathcal{|O|} }
\renewcommand{\b}[1]{\boldsymbol{\mathrm{#1}}}
\section{System Model and Problem Formulation}
\label{sec:problem_formulation}
\begin{figure}[t]
	\centering
	\includegraphics{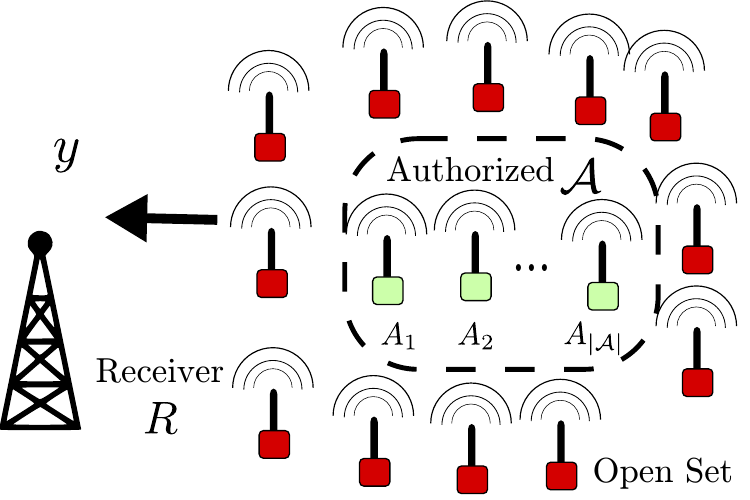}
	\caption{Signal $y$ is received by receiver $R$. We want to determine if it was sent by an authorized transmitter in the set $\mA$  or a new unseen transmitter.}
	\label{fig:problem_formulation}
\end{figure}

We consider a finite set of authorized transmitters given by  $\mA=\{A_1,A_2, \cdots,A_\mAc \}$ that are allowed to send data to a receiver~$R$, where $\mAc$ is the size of the set $\mA$. When some transmitter $T$ sends a set of symbols $\b{x}$, the signal received by $R$ is $f_T(\b{x},t)$. The function $f_T$ represents the  time variant RF fingerprint, which captures  the transmitter  hardware fingerprints and  wireless channel effects.  Since the channel depends on the environment surrounding the transmitters, it is more prone to temporal variation. 

 The authorization problem can be formulated as shown in Fig.~\ref{fig:problem_formulation}: receiver $R$ receives a signal $\b{y}$ from some transmitter $T$ and wants to determine whether the transmitter $T$ belongs to the authorized set or not without decoding $\b{y}$. This can be formulated as the following hypothesis test: 
\providecommand{\mPfa}{P_{FA}}
\providecommand{\mPd}{P_{D}}
\newcommand{\mHz}{\ensuremath{\mathcal{H}_0} }
\newcommand{\mHo}{\ensuremath{\mathcal{H}_1} }
\renewcommand{\b}[1]{\boldsymbol{\mathrm{#1}}}
\begin{align}
\label{prob:outlier_det}
\begin{split}
\mHz: & \ \b{y} = f_T(\b{x},t), T\in \mA \\
\mHo: & \ \b{y} = f_T(\b{x},t), T\notin \mA  
\end{split} \ \ \ \  \forall \ \  t 
\end{align}  
Here, \mHz corresponds to an authorized transmitter and  \mHo corresponds to an outlier. 

Additionally, in cases where each authorized transmitter has different privileges, we might be interested in classifying it within the authorized set, which can be formulated as finding  $\hat{A}$ that is most likely to have generated $\b{y}$, defined as
\begin{equation}
\label{prob:class}
\hat{A} = \underset{T}{\text{argmax}} \ \mathrm{Pr}(f_T(\b{x},t)=\b{y}|\b{y}) , \ \ \ \ T\in \mA, \ \ \ \forall \ t
\end{equation}

While the anomaly detection problem addresses only problem (\ref{prob:outlier_det}), the open set problem addreesse both (\ref{prob:outlier_det}) and (\ref{prob:class}). Since classification has been studied extensively in the literature, our main focus in this work is on the results of outlier detection when using either formulation.

To improve outlier detection,  we propose using an additional class of \textit{known} outlier transmitters $\mK=\{K_1,K_2,\cdots,K_\mKc\}$, where $\mK \not\subset \mA$. Samples from transmitters in $\mK$ will be used during training to assist the outlier detector to differentiate between authorized and non-authorized transmitters. But still, the evaluation of any outlier detector is done using a set of unknown outliers  $\mO$ such that $\mO \cap \mK = \emptyset$. In practice, samples from the set $\mK$ can be obtained by capturing data from a finite number of non-authorized transmitters. 

\section{Machine Learning Approaches}
\label{sec:ml_approach}
In this section, we  discuss machine learning approaches to address this problem. An  approach consists of  a neural network architecture, followed by an output processing stage to decide whether a signal is an outlier or not. For some approaches,  the sensitivity to outliers can be changed by modifying a threshold.

\begin{figure*}[t]
	\centering
	\subfloat[Disc \label{fig:net_disc}]{\includegraphics{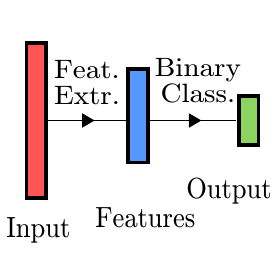}}  \hspace{8mm}
	\subfloat[DClass \label{fig:net_dclass}]{\includegraphics{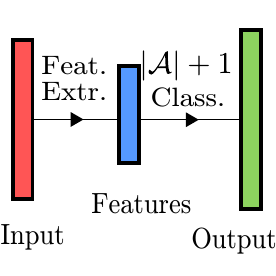}} \hspace{8mm}
	\subfloat[OvA \label{fig:net_ova}]{\includegraphics{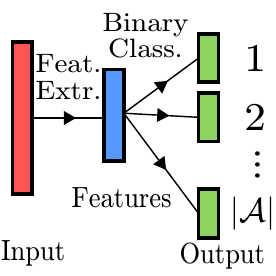}}  \hspace{8mm}
	\subfloat[AutoEnc \label{fig:net_autoenc}]{\includegraphics{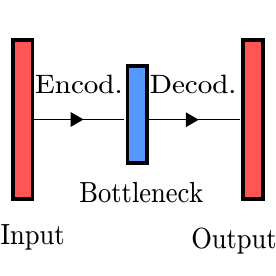}}   \hspace{8mm}
	\subfloat[OpenMax \label{fig:net_opmx}]{\includegraphics{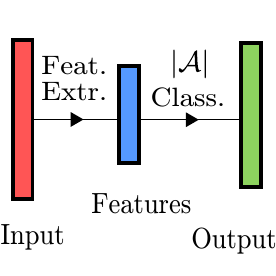}} 
	
	\caption{ High level architecture of the proposed methods. Autoencoder consists of an encoder and a decoder. The remaining ones consists of a feature extractor followed by one or more classifiers. }
	\label{fig:net}
	\vspace{-5mm}
\end{figure*}

\subsubsection{Discriminator (Disc)}
The most intuitive approach for outlier detection is to train a classifier that outputs a binary decision on whether the signal is an outlier or not. The main limitation of this method is its complete reliance on the known outlier set for training. 
 In terms of architecture, the  discriminator has as single scalar output $z$  as shown in Fig.~\ref{fig:net_disc}.  $z$ is generated by a sigmoid and takes a value between 0 and 1. The labels for authorized transmitters and outliers are $l=0$ and $l=1$, respectively.    A  threshold $\gamma$ is used to make a decision with $\mHo$  declared if $z>\gamma$; \mHz is declared otherwise. 

\subsubsection{Discriminating Classifier (DClass)}
To detect outliers and to classify among the authorized set, the most straightforward approach is to train a network with $\mAc+1$ outputs, where the additional class corresponds to outliers. This classifier is expected to perform better than Disc, as individually labelling transmitters should help it extract better features. To train this network, we also need known outliers similar to Disc.  A signal is classified as an outlier if the maximum activation corresponds to the last class; it is considered authorized otherwise. For this architecture, it is not straightforward to design an adjustable threshold as in Disc.

\subsubsection{One Vs All (OvA)}
A simple way to generalize Disc to perform multi-class classification is to train one discriminator network for each of the $\mAc$ authorized transmitters. However, this needs $\mAc$ networks the size Disc each with a feature extractor network. A better way to scale the discriminator is shown in Fig.~\ref{fig:net_ova}. In this approach, all $\mAc$ binary classifiers  share the same feature extractor similar to what was proposed in~\cite{shu_doc_2017}.  Unlike Disc, OvA  does not require a known set of outliers as long as $\mAc\geq 2$, since for  binary classifier $i$, signals from all transmitters $j\neq i$ are considered outliers.
The output of this network will be a vector $\b{z}$ of $\mAc$ real numbers such that $\b{0}\leq \b{z}\leq \b{1}$, where $\b{0}$ and $\b{1}$ are the vectors of all-zeros and all-ones, respectively. Following the notation in \cite{shu_doc_2017}, the labels for a sample from authorized transmitter $A_i$ will have   $l_i=1$ and  $l_j=0 ~\forall j\neq i$ and a known outlier, if used, will have $\boldsymbol{l}=\b{0}$.    

  The decision is based on $\mAc$  thresholds given by $\pmb{\gamma}$, where element $\gamma_i$ is the threshold for $z_i$. Each binary classifier $i$ declares a sample to belong to its class  if $z_i>\gamma_i$.  A signal is declared to be an outlier (corresponding to $\mHo$), if all discriminators declare the signal to be not within their class ($\b{z}\leq \pmb{\gamma}$), and to be within the authorized set (corresponding to $\mHz$) otherwise.

\subsubsection{ OpenMax (OpMx)}
\label{subsubsec:openmax}
OpenMax is a popular approach for openset recognition \cite{openmax_2015}. It consists of modifying a trained classifier with $\mAc$ softmax outputs using statistical analysis of activations to detect outliers.   The output activation vector $\b{v}$,  obtained prior to  softmax, is processed to generate a modified activations vector $\b{v}'$ having $\mAc+1$ outputs, with the additional output corresponding to outliers.  The modified activation vector is given by
\begin{equation}
\label{eq:openmax_mav}
	v'_i = 
	\begin{cases}
	v_i \omega_i, & i \in \{1,\cdots,\mAc\} \\
	\sum_{i=1}^{\mAc} v_i(1- \omega_i) & i= \mAc +1 
	\end{cases}
\end{equation} 
where $\omega_i$ represents our confidence in the membership of the given sample to class $i$.
The concept behind calculating the vector $\b{\omega}$ is that the activation vectors of samples belonging to the same class are similar, while those belonging to unseen classes are different from all classes. This is implemented by calculating the mean activation vector $\bar{\b{v}}_i$  for each class $i$  using the training set. The distance  $d_i(\b{v}) = \| \bar{\b{v}}_i - \b{v} \|$ represents the similarity of the sample generating vector $\b{v}$ to  class $i$. On the training set, for the correctly classified samples, the distance of each sample belonging to a class $i$ is calculated. Using extreme value theorem~\cite{kotz_extreme_2000}, the tail of the distribution is calculated by fitting the  $\tau$  samples with the largest $d_i$ to a Weibull distribution having parameters $(m_i,\eta_i)$. Then, for the $\alpha$ classes having the highest activations, $\b{\omega}_i$ is calculated by evaluating the probability of belonging to the tail of distribution of class~$i$ using
\begin{equation}
	\omega_i = 1 - R_\alpha(i) \times \text{WeibullCDF}(d_i(\b{v}),(m_i,\eta_i))
\end{equation}
where $R_\alpha(i) = \frac{\alpha -i}{\alpha}$ is a  calibrator with parameter $\alpha$ and   $\text{WeibullCDF}(x,(m,\eta)) =\exp \left(- \left(  x/\eta \right)^m \right)$, as explained in \cite{openmax_2015,yoshihashi_classification-reconstruction_2019}. After calculating the modified activation vector $\b{v}'$, uncertain outputs are rejected if the confidence is below some threshold $\epsilon$.  This is done by applying the softmax function to $\b{v}'$  to obtain the vector $\b{z}$. Then we calculate $i=\text{argmax}(\b{z})$;  the sample is considered an outlier if  $i=\mAc+1$ or $z_i< \epsilon$. Since OpenMax is based on a classifier, it does not benefit from known outliers in training.

The approach proposed for transmitter authorization in~\cite{gritsenko_finding_2019}   consists of modifying a classifier similar to OpenMax. However, their approach only uses the maximum value of the softmax output, while OpenMax uses the entire activation vector. Hence, it uses more information from the neural network output. %

\subsubsection{AutoEncoder (AutoEnc)}
Autoencoders are commonly used for anomaly detection~\cite{anomaly_detection_survey_2019}. They consist of an encoder mapping the data into a smaller dimension, the bottleneck, followed by a decoder trying to  reconstruct the original input. During training, autoencoders learn the distribution of the training data. When the autoencoder processes  anomalous data, it generates a higher  reconstruction error, which can be used to detect anomalies. During training, the objective of the autoencoder is to reduce the mean squared error, $\text{MSE} = \|\b{y}-\hat{\b{y}}\|$ where $\b{y}$  is the input and  $\hat{\b{y}}$ is the output of the autoencoder. A signal is considered an outlier if this error is bigger than some threshold   $\gamma$.

To summarize, the proposed approaches are either based on classifiers (Disc, DClass, OvA, OpMx) or uses signal reconstruction (AutoEnc). The classifier-based approaches share a neural network-based feature extractor and differ only in the output labels and the last layers activation function. Some of the  classifier-based approaches, beside outlier detection, classify the authorized signal among the set $\mA$. As for the known outlier set $\mK$, it is necessary for the training of some approaches, improves the performance of others, and cannot be used in some other approaches. Table~\ref{tbl:approach_features} provides a high level comparison of approaches. 
 
For the approaches which have an adjustable threshold,  a tight threshold would lead to signals from authorized transmitters being mistakenly rejected (high probability of false alarm $\mPfa$) and a loose threshold would fail to recognize many outlier signals (low probability of detection $\mPd$). The method of setting the thresholds along with any hyperparameters is discussed in Appendix \ref{ap:params}. 

\begin{table}[htbp]
	\renewcommand{\arraystretch}{1.5}
	\caption{Features of Approaches \label{tbl:approach_features}}
	\centering
	\begin{tabular}{|c|c|c|c|}
		\hline
		Approach & Works with $\mK$? & Adjustable Threshold & Classifies $\mA$ \\	\hline
		Disc & Necessary & Yes & No\\ 	\hline 
		DClass & Necessary & No & Yes\\	\hline 
		OvA & Yes & Yes & Yes\\	\hline 
		OpMx & No & No & Yes\\	\hline 
		AutoEnc & No & Yes & No\\	\hline 
	\end{tabular} 
\end{table}

\section{Dataset}
\label{sec:dataset}
The dataset was captured using off-the-shelf WiFi modules (Atheros 5212, 9220, and 9280) as transmitters and a software defined radio (USRP N210) as a receiver. The capture was performed in the Orbit testbed  grid \cite{orbit_2005}. Orbit testbed  grid consists of 400 nodes arranged in  a 20$\times$20 grid with a separation of one meter. The receiver was chosen near the center of the grid and 163 nodes surrounding it were used as transmitters.

The capture was done over Channel 11 which has a center frequency of 2462 MHz and a bandwidth of 20 MHz. %
Captures were taken at a rate of 25 Msps, over a duration of one second. After the IQ capture was complete, the packets were extracted using energy detection. The number of packets captured during this period for each transmitter varied according to the WiFi rate control and their total number  is over 300,000. %

 While it is possible to use the entire packet payload for training, this would make the data contained in each slice different. In our previous work \cite{hanna_icnc_2019}, we have shown that using the  slices with the same data leads to a better performance than using  slices containing random data. This was also verified in \cite{morin_transmitter_2019}.  Hence, from each captured WiFi packet, we used the first 256 IQ samples containing the preamble. The IQ samples were normalized  to have a unity average magnitude without any  further preprocessing. %
For transmitter authorization in \cite{gritsenko_finding_2019}, the entire packet payload was used for training and inference was done on slices of a packet, which are combined to obtain one decision per packet. While this method leads to having more data, it makes the learning task of the neural network  harder, as was previously demonstrated in~\cite{hanna_icnc_2019,morin_transmitter_2019}. Since in this work we only consider  preambles and due to the similarity of their approach to OpenMax, we don't consider their  approach in our evaluation.
 
As was pointed out in recent works \cite{cekic_robust_2020,al-shawabka_exposing_2020}, the fingerprints learned by neural networks can be dependent on the channel and not the transmitter. This causes significant degradation in the recognition performance if the testing data was captured on a different day than the training capture. To this end, using the previous setup, we made five data captures over 5 different days.  The data from the first capture is the one used in our previous work~\cite{hanna_spawc_2020}, and contained data from less transmitters. Also, it was made two months earlier than the remaining captures.  The data from the last day was kept exclusively for testing.

\section{Network Architectures}
\begin{figure}[t]
	\centering
	\includegraphics{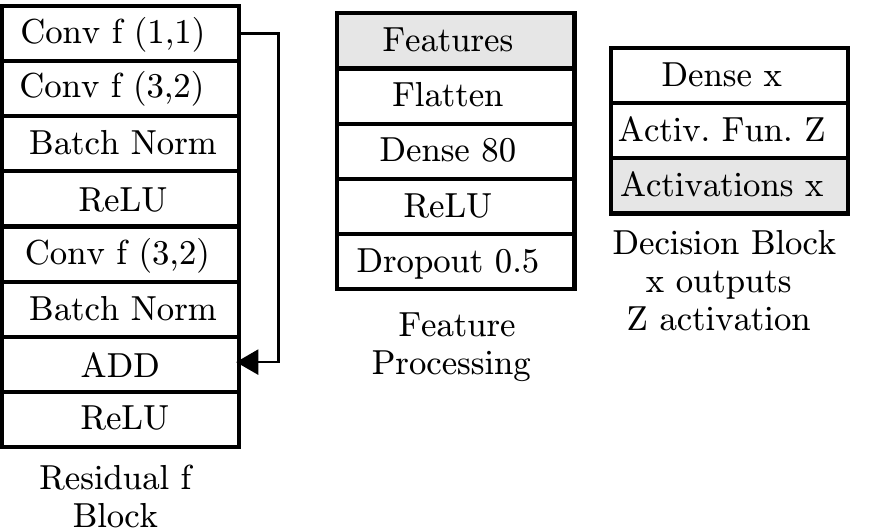}
	\caption{The residual, feature processing, and decision blocks used as  building blocks for the neural networks considered in this work.}
	\label{fig:net_layers_blocks}
\end{figure}
\begin{figure}[t]
	\centering
	\includegraphics{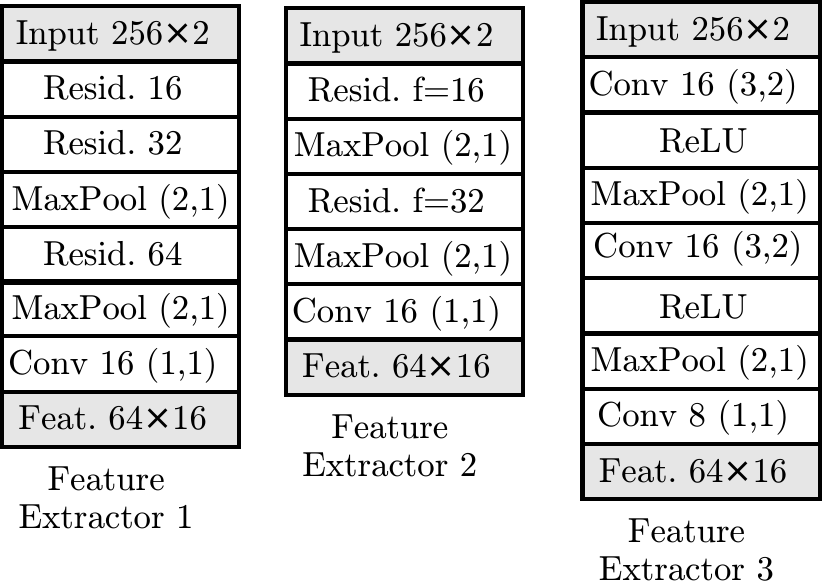}
	\caption{The three different architectures compared in this work. Feature extractor 2 was eventually selected.}
	\label{fig:net_layers_feat_extr}
\end{figure}
In this section, we  describe the architecture of the proposed networks.  Since the feature extractor is an essential component for Disc, DClass, OvA, and OpMx, we consider several alternatives for it and compare to the architecture used in our previous work \cite{hanna_spawc_2020}. All the architectures in this work are built using the blocks shown in Fig.~\ref{fig:net_layers_blocks}; the residual block which has $f$ filters~\cite{he_deep_2015}, the feature processing block, and the decision block using activation function $Z$ to generate a given number $x$  outputs.

\subsection{Architecture Comparison}

\begin{figure}[t]
	\centering
	\subfloat[OvA used in our prior work~\cite{hanna_spawc_2020} with a feature processing block per output. \label{fig:net_layers_ova_arch1}]{\includegraphics{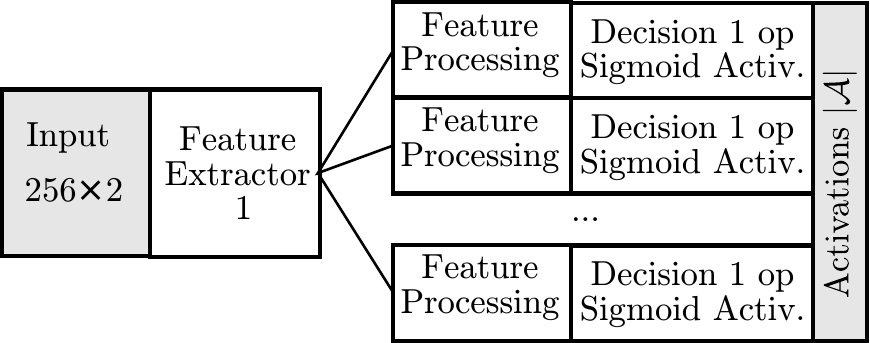}} \\
	\subfloat[OvA used in this work having a shared feature processing block. Several feature extractors were compared and eventually Feature Extractor 2 was used. \label{fig:net_layers_ova_arch2}]{\includegraphics{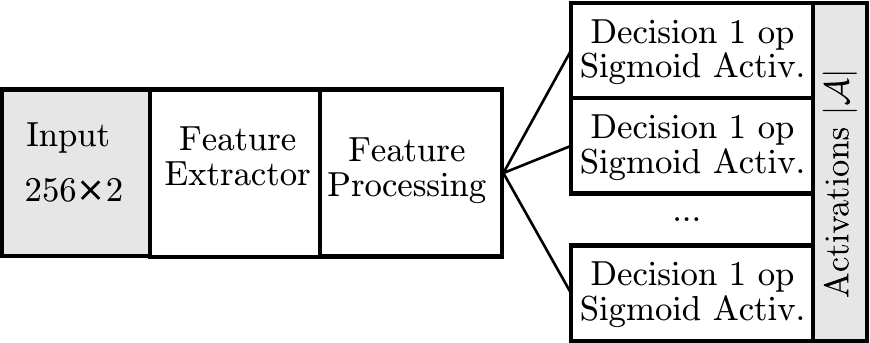}} 
	\caption{We considered two OvA architectures, the first having a feature processing block for each output, and the second using a common one.}
	\label{fig:net_layers_ova_arch}
\end{figure}
\begin{table}[t]
	\renewcommand{\arraystretch}{1.5}
	\caption{OvA Network Sizes \label{tbl:net_ova_size}}
	\centering
	\begin{tabular}{|c|c|c|}
		\hline
		OvA & Description & \# of  params. \\	\hline
		1 & Feat. Ext. 1 + 10 Feat. Process. & 890,170 \\ 	\hline %
		2 & Feat. Ext. 1 + Common Feat. Process. & 152,170 \\	\hline %
		3 & Feat. Ext. 2 + Common Feat. Process.  & 99,754 \\	\hline %
		4 & Feat. Ext. 3 + Common Feat. Process. & 46,226 \\	\hline %
	\end{tabular} 
\end{table}
\begin{figure}[t]
	\centering
	\subfloat[Classification Accuracy \label{fig:arch_class}]{ \includegraphics{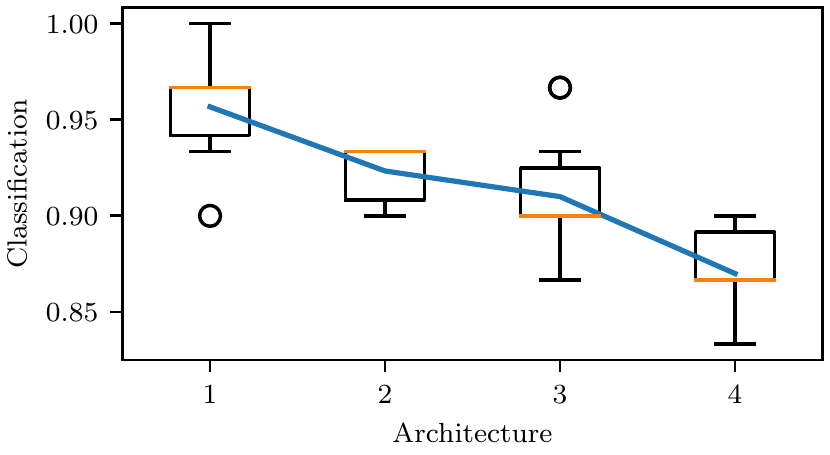}} \\
	\subfloat[Outlier Detection Accuracy \label{fig:arch_acc}] {\includegraphics{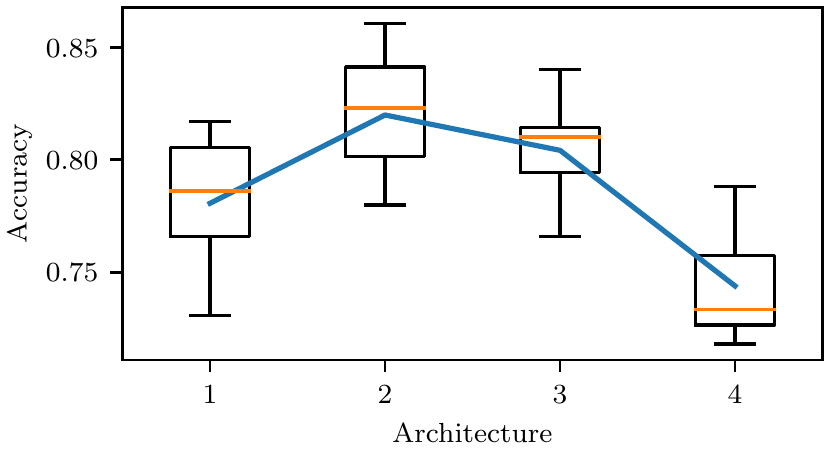}} 
	\caption{Average performance of different architectures of OvA shown as a blue solid line. Box plots represent the variation due to training 10 repetitions of the same network using the same data. Improvement in classification is not necessarily correlated with improvement in outlier detection.}
	\label{fig:arch}
\end{figure}

For choosing the best architecture and contrasting to the architectures in our previous work, we use OvA as a benchmark using data from 10 authorized transmitters ($\mAc=10$) collected on the same 4 days and no known outliers. Details for the network training is deferred for later in the paper.

We consider 4 architectures of OvA. OvA 1 (from~\cite{hanna_spawc_2020}) which has $\mAc$ feature processing block and OvA 2, which uses the same feature extractor 1 with a common feature processing blocks as shown in Fig.~\ref{fig:net_layers_ova_arch}. Moving from OvA 2 to OvA 3 and OvA 4, we use smaller feature extractors 1,2, and 3 respectively which are described  in Fig.~\ref{fig:net_layers_feat_extr}.  The summary of each architecture and the number of trainable parameters are shown in Table~\ref{tbl:net_ova_size} from which we see that from OvA 1 to OvA 4 the network gets smaller.

Due to the random initialization of the weights, along with randomness in batch division during training, the same network trained using the same data can give different results. To have confidence on the significance of our results, each network is trained from scratch using the same data for ten repetitions and the statistics of the results are shown.  Fig~\ref{fig:arch_class} shows the classification results of the authorized nodes.  The larger networks perform better, as expected; by having more learning capacity, the networks are better at distinguishing between transmitters. But looking at the accuracy of outlier detection in Fig.~\ref{fig:arch_acc} for OvA 1 and 2, we see a rather surprising result; the largest network actually performs worse than some of the smaller networks. Since we want the network to generalize to new transmitters, we want each binary classifier of OvA to learn only the characteristics of its designated transmitter, while rejecting any other transmitter. But once the learning capacity of the per transmitter branches increases beyond a certain point, %
it starts to learn the characteristics of the remaining transmitters. Although this improves classification and minimizes training and validation loss, it does not generalize well to outlier detection.

As we decrease the capacity of the common feature extractor in OvA 2 to 4, the performance of both classification and outlier detection degrades. Since this shared block extracts features and does not make a decision, the more learning capacity it has, the better the results.  Although feature extractor 1 performs about 1\% better, as a design choice, we  use feature extractor 2  for the rest of this work  because of its smaller network size. For the remaining of this work, we use OvA 3, having the common feature processing architecture.

\subsection{Architecture Description}

\begin{figure}[t]
	\centering
	\includegraphics{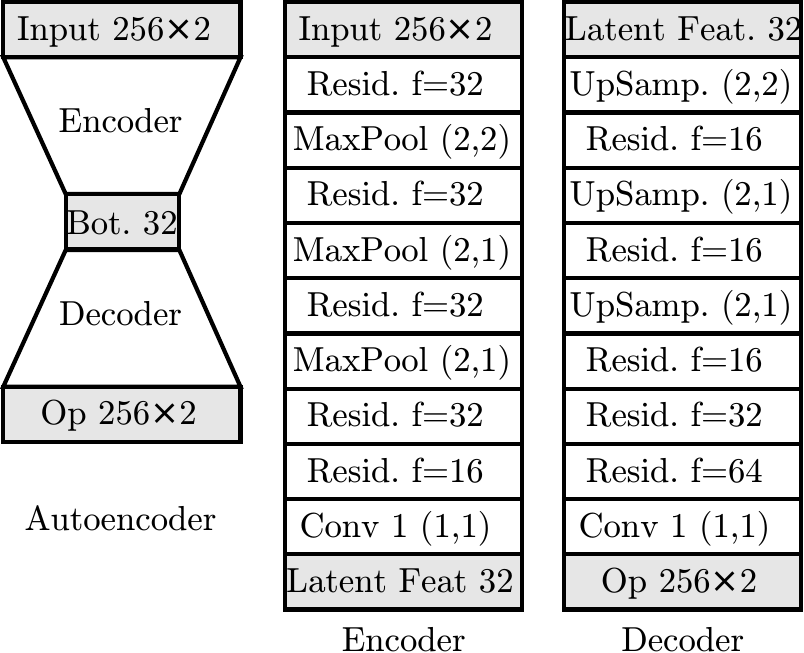}
	\caption{The architecture of the autoencoder.}
	\label{fig:net_layers_autoenc}
\end{figure}

The architectures for Disc, DClass, and OpMx consist of Feature Extractor 2 followed by a feature processing block and a decision block with 1 sigmoid, $\mAc+1$ softmax, and $\mAc$ softmax respectively.  Disc was provided with a larger feature processing network having an additional Dense network with 100 neurons after the flattening to be comparable in size to the other networks. The autoencoder architecture is shown in Fig.~\ref{fig:net_layers_autoenc}. It consists of encoder with a bottleneck consisting of 32 samples followed by a decoder which reconstructs the signal. 

 The number of trainable parameters of each network is shown in Table~\ref{tbl:net_size}.  The network sizes of OvA, DClass, and OpMx scale with $\mAc$\footnote{In our previous work \cite{hanna_spawc_2020}, since we repeated the feature processing block for OvA, the network size increased with $\mAc$ at a much higher rate}.  AutoEncoder and Disc have a fixed number of parameters for any value of $\mAc$.
  Notice  that OvA and OpMx have an identical number of parameters while DClass differs by only an additional 81 parameter.  
  
  From an  architecture perspective, Disc, OvA, OpMx, and DClass are very similar. They share the same feature extractor and feature processing blocks, which constitute most of their neural network.   The difference between these methods  come from the type of activation function, data labeling, and post processing performed. These differences lead to conceptual differences in the approach as we have discussed, leading to different characteristics as we summarized in Table~\ref{tbl:approach_features}, and will lead to significantly different performance.

\begin{table}[htbp]
	\renewcommand{\arraystretch}{1.5}
	\caption{Network Size \label{tbl:net_size}}
	\centering
	\begin{tabular}{|c|c|c|}
		\hline
		Net & \# of trainable prams. \\	\hline
		OvA &  98944 + 81 $\mAc$ \\ 	\hline %
		Disc  & 127,605  \\	\hline
		DClass  & 99,025  + 81 $\mAc$ \\	\hline
		AutoEnc & 109,362 \\	\hline
		OpMx & 98944 + 81 $\mAc$  \\	\hline
	\end{tabular} 
\end{table}

\section{Evaluation Procedure}
\label{sec:results}
\begin{figure}[t]
	\centering
	\includegraphics{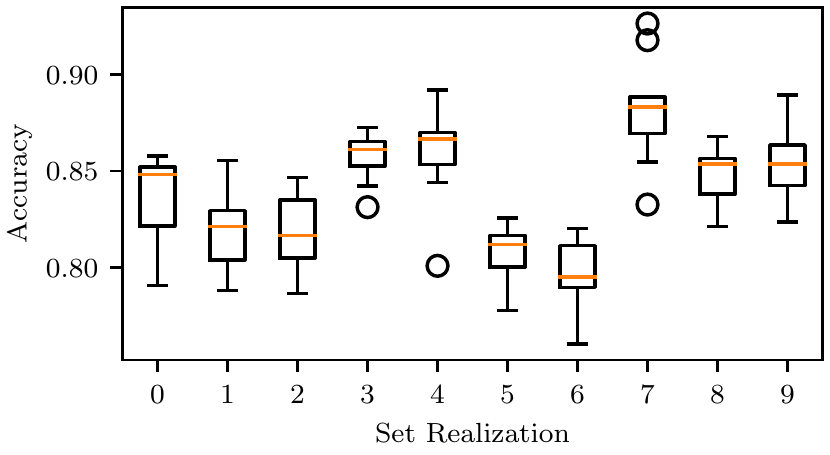}
	\caption{ Box plot of the outlier detection accuracy of OvA is shown for several realizations of the sets. For each set realization, a network is trained for 10 repetitions. The center line represents the median, the box represents the first and third quartiles, and the whiskers represent the range, except for outlier points which are represented as circles. }
	\label{fig:sets_acc}
\end{figure}

In this section, we describe the evaluation procedures used through out this work. We discuss the evaluation metrics used, steps to avoid the dependence on the specific division of the dataset to different sets ($\mA$, $\mK$, $\mO$), and the way we evaluate how the approach generalizes across time.   

As stated earlier, for many of these networks, there is a threshold which defines the trade-off between detection and false alarm. This tradeof is typically represented by a Receiver Operating Characteristic  (ROC) curve.  To compactly visualize the results, we consider the area under the ROC curve (AUC), which measures performance in a manner independent of threshold~\cite{sun_classification_2009}.  Still, when the network is deployed, we have to choose a threshold. Given a specific threshold-set as discussed in Appendix \ref{ap:params}, we calculate the accuracy of correctly identifying outliers  over a balanced test set, such that random guessing would yield a 50\% accuracy. In that case, the accuracy is the average of the performance on the authorized samples given by $1-\mPfa$ , and the outliers given by $\mPd$. Classification accuracy  results  included for the authorized samples  are evaluated for a balanced version of the test set having the same number of the samples from each authorized transmitter, where any trivial or random guess would yield an accuracy of $1/{\mAc}$\%. 
Note that as stated earlier, not all methods have an adjustable threshold (as summarized in  Table~\ref{tbl:approach_features}) and hence won't have a corresponding AUC result.

Unlike with classification, where typically all transmitters available are used, outlier detection involves dividing the transmitters into sets of authorized transmitters, outliers, and possibly known outliers. Since RF fingerprints are random, some transmitters are more similar than others, and a comprehensive evaluation cannot be done using only one realization of the sets. This adds another source of variability to our results, besides the inherent randomness in training neural networks. To demonstrate this, we train OvA using 10 authorized nodes and evaluate it using 63 outlier nodes picked randomly from the 163 transmitters. Ten random  realizations of these sets are compared and for each we train 10 repetitions. The results are shown as a box plot in Fig.~\ref{fig:sets_acc}, from which we can see up to 9\% difference in the median due to the different realizations of the sets. This is more significant than the difference between the first and third quartiles due to training randomness which did not exceed 3\%. Based on these findings, our evaluation considers multiple realization of the sets while only considering one repetition.

As for assessing the ability of our network to generalize through time, we create two test sets: a same-day test set which was captured on the same days as the training set and the different-day test set, which was captured on a different day than the training data.

Based on the previous considerations, we describe our evaluation procedure. For certain values of $\mAc$, $\mKc$, and $\mOc$,   we randomly partition the dataset to $\mA$, $\mK$, and $\mO$ to form 10 realizations of $\{\mA, \mK, \mO\}$. All approaches are evaluated using the same 10 realizations  and the results are shown in terms of mean and standard deviation. For  the  training data and the same-day test data, we use only samples from the captures made on the first four days, while the last day capture is entirely left for different day testing.

Given a realization of $\mA$, $\mK,$ and $\mO$, for training and validation, we use 70\% of the samples belonging to $\mA$, and all the samples belonging  $\mK$, from the same day data.  The  combination of this data is split into 80\% for training and 20\% for validation. The same day test set contains all samples from $\mO$  and the remaining 30\% of $\mA$. For different realizations of the sets, the dataset can get highly imbalanced. To avoid degenerate solutions, where the network always predicts the class with the majority of samples, the training loss is weighted based on class frequency. 
As for the different day test set, it is obtained by combining all samples from $\mA$ and $\mO$ from the last day capture.

The training used 10 epochs using the ADAM optimizer with a learning rate of 0.001. The weights corresponding to the epoch which produced the lowest validation loss are kept. Data was first normalized, then augmented by adding noise with a variance of 0.01 and applying a uniform random phase shift. Cross-Entropy was used as the loss function for Disc, DClass, OvA, and OpMx with classes weighted depending on the number of samples of each class. AutoEnc used MSE loss.

\section{Transmitter Set Sizes Evaluation}
In this section, we explore the effect of changing the size of the required authorized set $\mA$, and evaluate the effect of having a known outlier set  $\mK$ and its size on the ability of the network to distinguish authorized signals from outliers. Throughout this Section, we used $\mOc=63$ for the evaluation. %

\subsection{Authorized set}
\begin{figure*}[t]
	\centering
	\subfloat[AUC -- Same Days \label{fig:n_auth_auc}] {\includegraphics{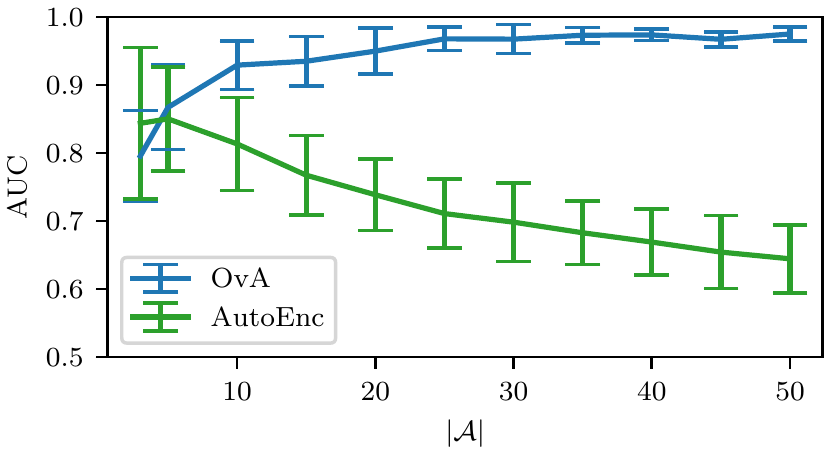}} 
	\subfloat[AUC-- Different Day \label{fig:n_auth_auc_t}] {\includegraphics{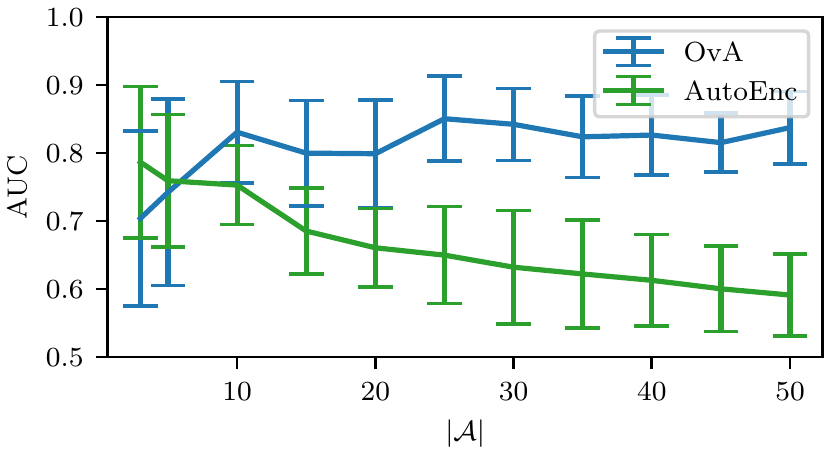}} \\
	\subfloat[Accuracy -- Same Days \label{fig:n_auth_acc}]{ \includegraphics{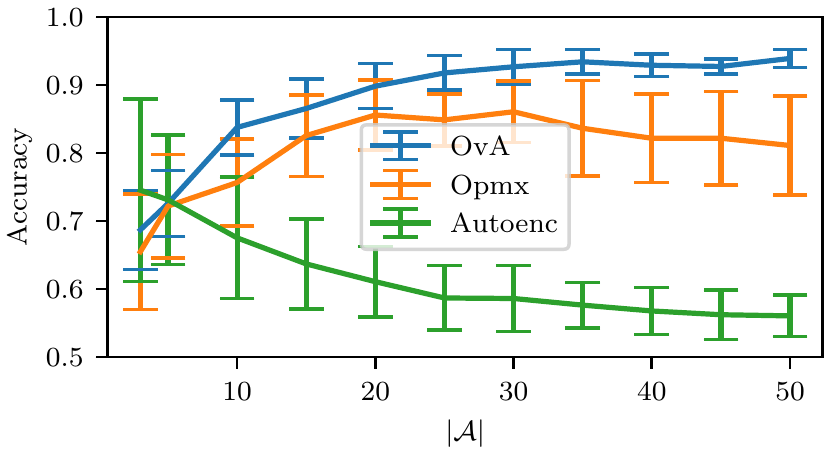}} 
	\subfloat[Accuracy -- Different Day \label{fig:n_auth_acc_t}]{ \includegraphics{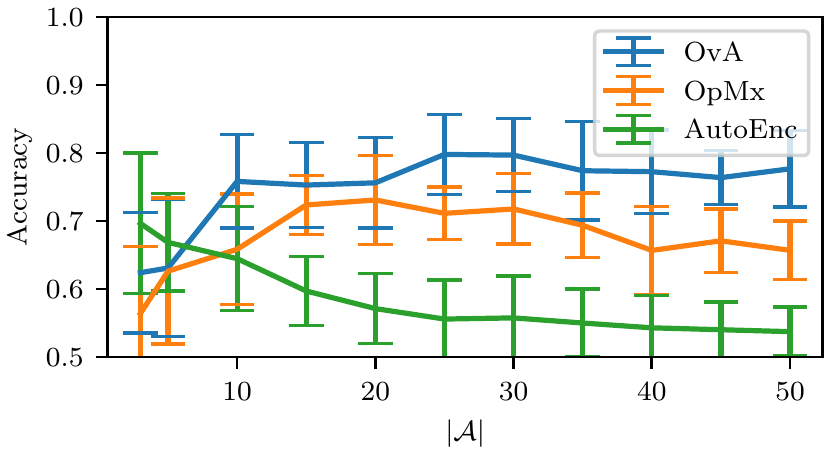}} \\
	\caption{Average outlier detection performance of several approaches as we change  $\mAc$. Error bars represent the standard deviation for different realizations of the sets.}
	\label{fig:n_auth}
\end{figure*}

\begin{figure*}[t]
	\centering
		\subfloat[Probability of False Alarm \label{fig:n_auth_pfa}]{ \includegraphics{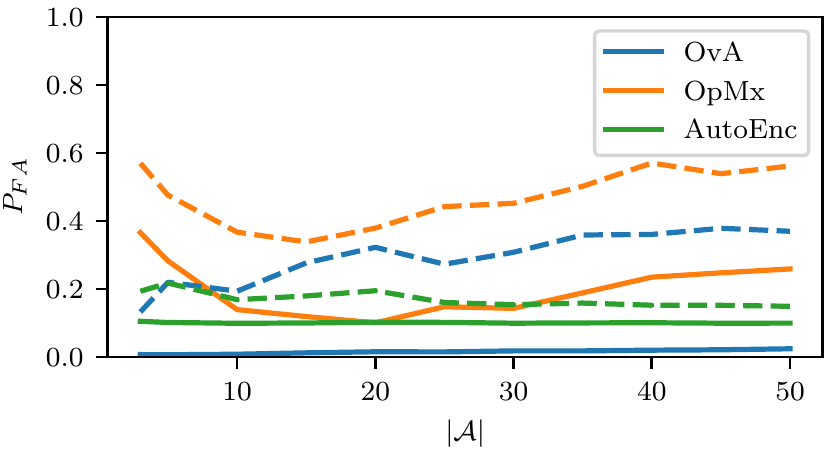}} 
		\subfloat[Probability of Detection \label{fig:n_auth_pd}]{ \includegraphics{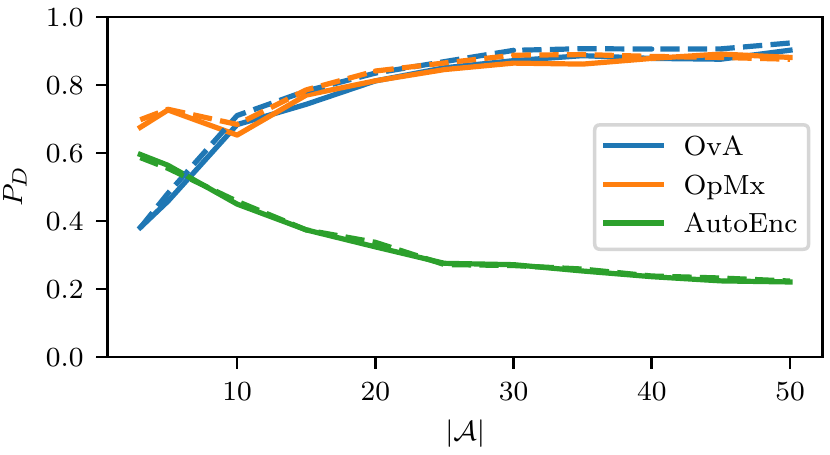}}
			\caption{Average outlier detection performance of several approaches as we change  $\mAc$. Error bars are omitted for clarity. Solid lines represent same days test, and dashed line represent different day test.}
			\label{fig:n_auth_p}
\end{figure*}
\begin{figure*}[t]
	\centering
	\subfloat[Accuracy -- Same Days  \label{fig:n_kn_out_acc}]{ \includegraphics{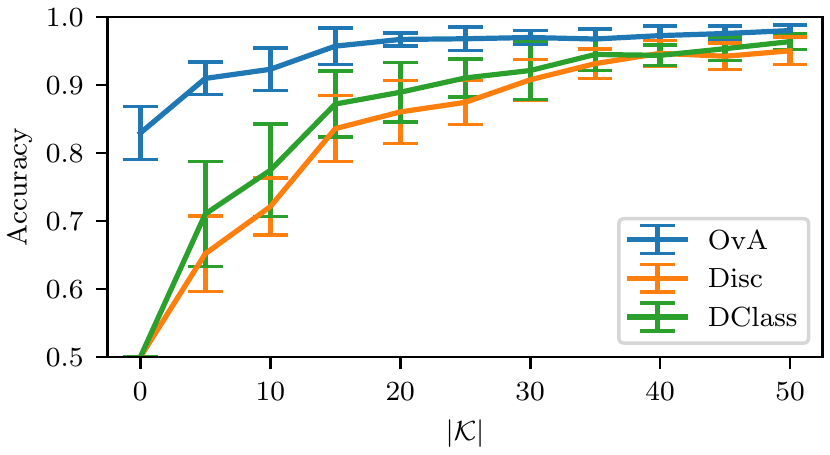}}
	\subfloat[Accuracy -- Different Day \label{fig:n_kn_out_acc_t}]{ \includegraphics{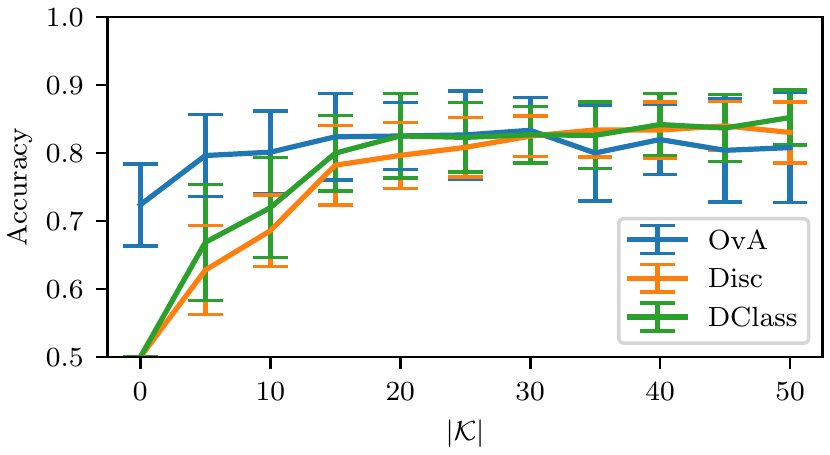}}
	\caption{Average outlier detection performance of several approaches as we change  $\mKc$. Error bars represent the standard deviation for different realizations of the sets.}
	\label{fig:n_kn_out_auth}
	\vspace{-4mm}
\end{figure*}
\begin{figure*}[t]
	\centering
	\subfloat[Probability of False Alarm \label{fig:n_kn_out_pfa}]{ \includegraphics{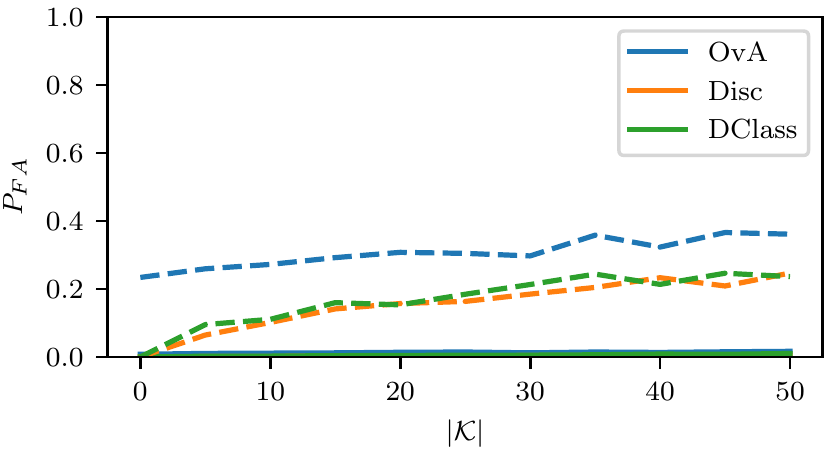}} 
	\subfloat[Probability of Detection \label{fig:n_kn_out_pd}]{ \includegraphics{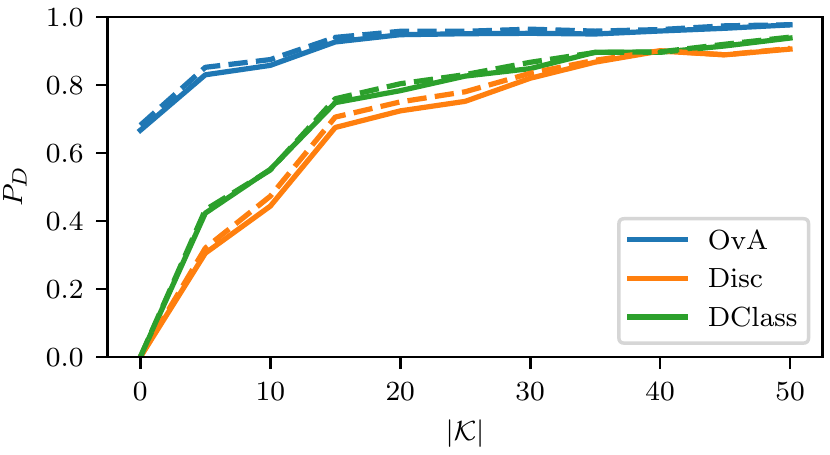}}
	\caption{Average outlier detection performance of several approaches as we change  $\mKc$. Error bars are omitted for clarity. Solid lines represent same days test, and dashed line represent different day test.}
	\label{fig:n_kn_out_p}
\end{figure*}

We start the evaluation by considering no known outliers, i.e, $\mKc=0$.  We want to know how large the set $\mA$ has to be for good outlier detection and what performance can be achieved thereof.  Results are shown for AUC and accuracy in Fig.~\ref{fig:n_auth_auc} and Fig.~\ref{fig:n_auth_acc} for the same-day test. 
For OvA, we see that as we increase the number of authorized nodes, the average AUC increases and its standard deviation decreases, showing less dependence on the set realization.  The accuracy, shown in Fig.~\ref{fig:n_auth_acc}, follows the same trend, and we are able to achieve accuracies above 90\% on the average when $\mAc$ is more than 20. The reason behind this pattern is that as $\mAc$ increases, each binary classifier has more signals from other transmitters, helping it learn better its designated transmitter without memorizing others, leading to better generalization. This interpretation is supported by the observation that the improvement in accuracy is due to the decrease in $\mPd$  for the same $\mPfa$ as shown in Fig.~\ref{fig:n_auth_pd} and Fig.~\ref{fig:n_auth_pfa}, respectively.

As for autoencoders, the trend seems to be reversed in Fig.~\ref{fig:n_auth_auc} and Fig.~\ref{fig:n_auth_acc}. As $\mAc$ increases, both the accuracy and AUC decrease. Autoencoders generate a compressed representation of their input by memorizing their distribution. For small $\mAc$, this is the distribution of the authorized transmitters.  As $\mAc$ increases, they learn the distribution of signals in general, independent of the transmitter. Hence, they reconstruct signals for unknown transmitters with low MSE and fail to detect them. This is verified by looking at the decreasing $\mPd$ curve for AutoEnc in Fig.~\ref{fig:n_auth_pd}, while $\mPfa$ is almost constant in Fig.~\ref{fig:n_auth_pfa}. This trend coincides with our visualization in Fig.~\ref{fig:ex_anomaly}.

For OpMx, the accuracy increases until $\mAc=20$ and then slightly decreases. This fluctuation is mostly attributed to $\mPfa$, as shown in Fig.~\ref{fig:n_auth_pfa}. The results of OpMx depend on the value of the activation vectors (AV) with respect to tail distributions  and the uncertainty threshold $\epsilon$. The key is understanding that the modified activations $\b{v}'$ calculated using~(\ref{eq:openmax_mav}) reduces the AV of the top $\alpha$ classes. After calculating the output $\b{z}=\mathrm{softmax}(\b{v}')$, the maximum $z_i=\mathrm{max}\{\b{z}\}$    is thresholded using $\epsilon$.  For small $\mAc$, classification for authorized is highly confident, leading to AV belonging to the tails of other classes, pushing $z_i$ below $\epsilon$, and leading to false alarms. As $\mAc$ becomes larger, this confidence decreases, leading to an improvement in $\mPfa$. However, the similarity between AVs of different classes decreases, pushing them to the tail  of other distributions as $\mAc$ increases beyond a certain point, leading to an increase in $\mPfa$.

In comparison, we see that for small $\mAc$, namely $\mAc = 3$, AutoEnc gives the highest accuracy on average because it is able to capture the distributions of small number of transmitters. At $\mAc = 5$, all methods are equally as good. As $\mAc$ increases, OvA gives the best performance because each branch uses all the data to learn its transmitter without memorizing the other transmitters. %

In Fig.~\ref{fig:n_auth_acc_t}, we plot the  accuracy for a different day test. The plots follows the same trends, but the accuracy of OvA and OpMx drop by about 15\% while AutoEnc drops by only 5\%. The reason behind this drop is clearer by inspecting the $\mPfa$ and $\mPd$ separately shown as dashed lines in Fig.~\ref{fig:n_auth_p}. From Fig~\ref{fig:n_auth_pd}, we see that the performance in detecting the new transmitters is almost unaffected.  The drop in accuracy is a result of the failure to identify authorized transmitters as shown in the $\mPfa$ curves in Fig~\ref{fig:n_auth_pfa}. This is reasonable since any change on unseen transmitters should not have any effect, unlike changes in the learned transmitters. For OvA, we see that as we increase $\mAc$, $\mPfa$ increases, since identifying transmitters from data captures on different days becomes even more challenging as we increase the number of transmitters.  The smaller  gap on different day test using AutoEnc is explained by the encoders ability to learn more general features of the signal compared to OvA and OpMx, which leads to an overall smaller drop in accuracy, which also causes the lower $\mPd$.

\subsection{Known set}
We expect that seeing more known outliers would help the network  differentiate the authorized transmitters from the outliers. To show this, we evaluate the performance of the  approaches that support having $\mK$ as input, as a function of $\mKc$ given $\mAc=10$.  The  accuracy curves are shown in Fig.~\ref{fig:n_kn_out_acc}. As stated earlier, at $\mKc=0$, DClass and Disc don't have any outlier samples for training and predict everything as authorized. 
From Fig.~\ref{fig:n_kn_out_acc}, we see that the accuracy of all methods improve as we increase the number of known outliers.  We also note that OvA is performing noticeably better than the others. This can be understood by realizing that in OvA, each binary classifier sees more samples to reject, the known outliers and the samples from other authorized transmitters. Thus, it is able to isolate its class better. DClass and Disc, on the other hand, only learn to reject samples from $\mK$. This is further supported by looking at the curves for $\mPfa$ and $\mPd$ shown in Fig.~\ref{fig:n_kn_out_pfa} and Fig.~\ref{fig:n_kn_out_pd}, where we see that the accuracy improvement is due to the probability of detection.
DClass slightly outperforms Disc because  the labels of $\mA$ help it extract better features compared to Disc. So it can be concluded that even if we are not interested in classifying among the nodes in $\mA$, including these labels in training improves the outlier detection performance.
Fig.~\ref{fig:n_kn_out_acc_t} shows the  accuracy curves when using the data from a different day for testing. Again, we see the same trend from the previous section, both AUC and accuracy drop by about 15\% for all methods due to the degradation of~$\mPfa$.%

\section{Dataset Training Days Evaluation}
\begin{figure*}[t]
	\centering
	\subfloat[Probability of False Alarm \label{fig:n_datasets_pfa}]{ \includegraphics{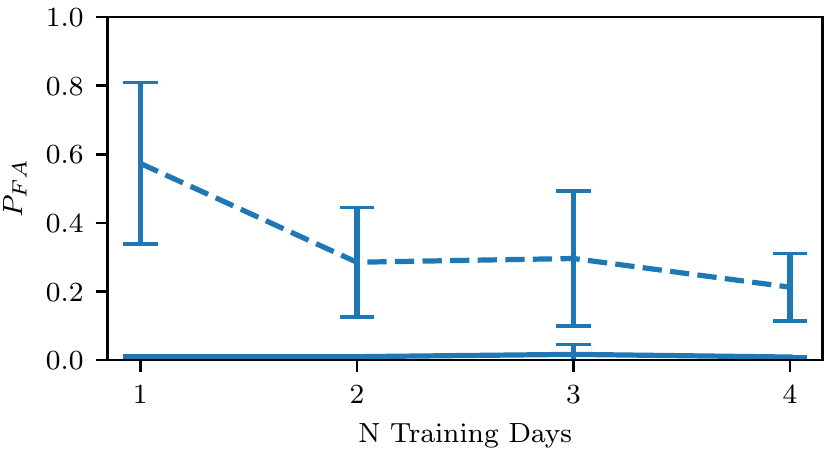}} 
	\subfloat[Probability of Detection \label{fig:n_datasets_pd}]{ \includegraphics{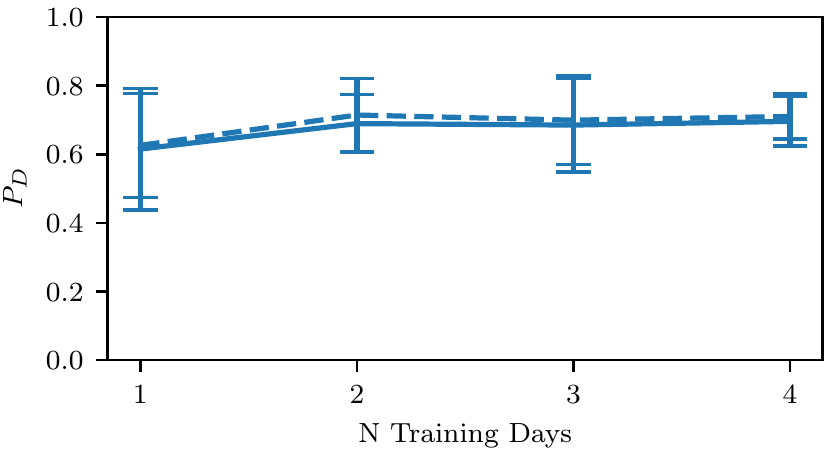}}
	\caption{Average outlier detection performance against the number of training days. Error bars represent the standard deviation for different set realizations.}
	\label{fig:n_datasets_p}
\end{figure*}
In this section, we evaluate the effect of the dataset construction on the ability of the proposed approaches to generalize over time. While developing methods to counter temporal variation in RF fingerprints is not the main focus of the paper, we study its effect on transmitter authorization.  For our evaluation, we only consider the OvA architecture with $\mAc=10$ and $\mKc=0$. We built four datasets, where dataset $i$ contains the data captures on dates prior to and including day $i$. The network was trained according to the same procedures discussed earlier and the results are shown in Fig.~\ref{fig:n_datasets_p}. From Fig.~\ref{fig:n_datasets_pfa}, we see that as the number of days included in training increases, $\mPfa$ decreases. On the other hand, $\mPd$ is almost unaffected.  The larger improvement from 1 day capture to two day captures is explained by the fact that the capture on day 1 was two months earlier than the remaining captures. During this long period, the transmitter fingerprints changed more significantly. The remaining captures were done on consecutive days, during which fingerprints had less severe changes, and hence smaller improvements to~$\mPfa$.  Hence, a simple way to improve the robustness against temporal variation is to collect data from the authorized transmitters over an extended period of time. Still, more sophisticated approaches are needed to close the gap between same day and different day testing.

\section{Summary}
\begin{figure}[h]
	\centering
	\includegraphics{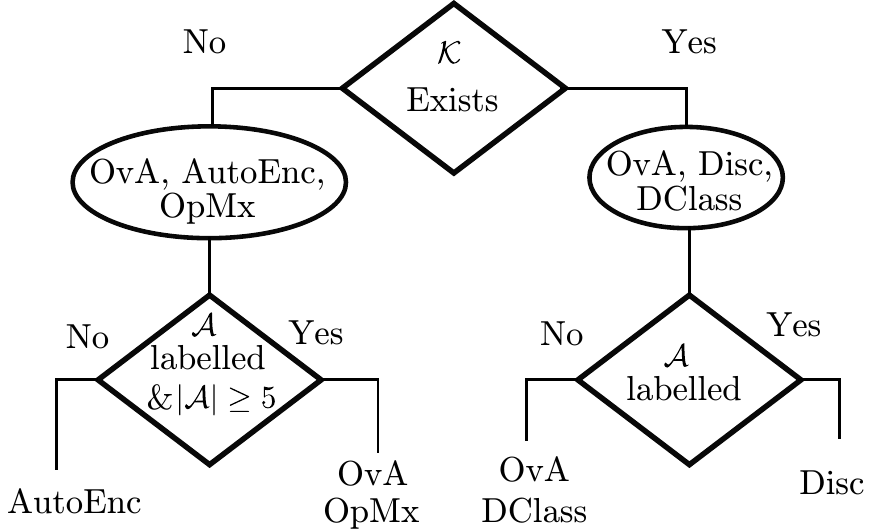}
	\caption{A tree diagram summarizing the feasible architectures as a function of the dataset. The networks are ordered so that the one yielding the better performance comes first.}
	\label{fig:approach_tree}
\end{figure}
The results we obtained can be summarized as follows.
\textit{Regarding the dataset:} It is better to label the authorized transmitters even if we are not interested in classifying among them, as it enables us to use openset methods (OvA, OpMx, DClass) which outperform the anomaly detection methods (Disc, AutoEnc) in many cases. Furthermore, the data for authorized transmitters should ideally be collected over a span of multiple days; as we have demonstrated, the drop in outlier detection accuracy is due to misclassification of authorized signals, i.e $\mPfa$, and can be reduced by collecting the data over multiple days.  

\textit{Regarding the approach:} The dataset structure; whether it is labeled or not, and the sizes of $\mA$ and $\mK$, determines which approaches are feasible or are better. If we have no data from known outliers and no labels, our only option is AutoEnc.  Without known outliers and with the availability of labels, for $\mAc\leq 5$, ignoring the labels and using AutoEnc is a better choice driven by its superior performance for small $\mAc$; if $\mAc > 5$, OvA gives the best performance. If we have a pre-trained classifier, then  OpMx is the best option. If we have $\mK$, we can use OvA, Disc, and DClass. Without labels, we are limited to Disc. If we have labels, OvA is typically better than DClass with tunable thresholds. This is summarized in Fig.~\ref{fig:approach_tree}.

\textit{Regarding the network architecture:} We have shown that if we have labeled data, the performance on classifying the transmitters does not necessarily correlate with the performance on outlier detection. A known outlier set is recommended to optimize the architecture.

Eventually, if we are able to collect the data as we wish (having labels, data for authorized transmitters captured over multiple days, and with data for known outliers) the best approach is OvA. When using 10 authorized transmitters and 50 known outliers, it  yielded an outlier detection accuracy of 98\% on the same day test set and 80\% on the different day test set.

\section{Conclusion}
\label{sec:conclusion}
In this paper, we have considered the problem of transmitter authorization using RF fingerprints captured from raw IQ samples. Since this problem has been scarcely investigated in the wireless domain, we performed a comprehensive evaluation of the most prominent machine learning approaches from the openset recognition and the anomaly detection literature, as applied to our problem definition. The dependence of the evaluation results on the choice of transmitters was demonstrated and a simple strategy was proposed to reduce it. We have also shown that the performance of a given neural network model on closed set classification is not an indicator of its performance in outlier detection, indicating the need for architectures specifically designed for this problem. Also, we demonstrated that minor change in network architecture and  data labeling can lead to a significantly different approaches.  Using a known outlier set was proposed and was shown to improve the outlier detection accuracy.  While classification based OvA gives the highest accuracy in most cases, it is outperformed by reconstruction based AutoEnc for small number of authorized. This opens the door for hybrid approaches combining  classification and reconstruction. We also pointed out that the temporal variation of fingerprints is  an open problem for transmitter authorization.

\appendices
\section{Parameter Selection}
\label{ap:params}
 For each approach, we describe how the ROC curve is calculated.  We also state how a specific threshold is chosen to calculate the outlier detection accuracy along with choices of  hyperparameters.  Since the evaluation is to be done over multiple realizations, manual tuning is not possible and we provide a systematic way to set these values.
\subsubsection{Discriminator (Disc)} In Disc, we only have one threshold to make a decision. Ideally, we want the threshold to be as low as possible without falsely rejecting authorized transmitters. This can be done by adapting the threshold to tightly fit the predictions of authorized signals in the training set. We follow the approach proposed in \cite{shu_doc_2017}, where the predicted output of the sigmoid for the correctly classified authorized training samples $\bar{z}_0$ (having labels equal to 0) is concatenated with its negative $-\bar{z}_0$ (to make the distribution symmetric around zero) and fit to a Gaussian distribution having mean 0. The standard deviation $\sigma$ of these samples is calculated and a threshold of $3\sigma$ would allow the majority of authorized transmitters to be accepted. To deal with degenerate cases having large standard deviation, the threshold is set to  $\gamma= \min(0.5,  3\sigma)$ in practice.  As for obtaining the ROC curve, the value of $\gamma$ is scanned from 0 to 1.
\subsubsection{One Vs All (OvA)}
OvA  has $\mAc$ thresholds given by $\pmb{\gamma}$. While it is possible to use one common threshold, we use multiple thresholds designed according to the same method of  Gaussian fitting used in Disc to calculate the accuracy as it yields  better results. As for obtaining the ROC curve, to be able to obtain a full ROC curve,  we consider one single threshold $\gamma$  scanned from 0 to 1 such that $\pmb{\gamma}=\gamma\b{1}$. 
\subsubsection{OpenMax (OpMx)}
As for the parameters, the tail size used to calculate the  Weibull distribution is $\tau=10$,  $\alpha= \min( \left \lfloor{\mAc/3}\right \rfloor  ,5) $, and $\epsilon$ was chosen to be the 95\% quantile of the maximum activation in the training data. These values were either adapted from~\cite{openmax_2015} or obtained empirically. 
\subsubsection{AutoEncoder (AutoEnc)}: We chose $\gamma$  to be the 90\% quantile of the mean squared error of the training data. The  ROC curves are obtained by scanning the value of $\gamma$ from 0 to the max MSE.

\ifCLASSOPTIONcaptionsoff
  \newpage
\fi

\bibliographystyle{IEEEtran}
\bibliography{references}

\begin{thebibliography}{10}
\providecommand{\url}[1]{#1}
\csname url@samestyle\endcsname
\providecommand{\newblock}{\relax}
\providecommand{\bibinfo}[2]{#2}
\providecommand{\BIBentrySTDinterwordspacing}{\spaceskip=0pt\relax}
\providecommand{\BIBentryALTinterwordstretchfactor}{4}
\providecommand{\BIBentryALTinterwordspacing}{\spaceskip=\fontdimen2\font plus
\BIBentryALTinterwordstretchfactor\fontdimen3\font minus
  \fontdimen4\font\relax}
\providecommand{\BIBforeignlanguage}[2]{{%
\expandafter\ifx\csname l@#1\endcsname\relax
\typeout{** WARNING: IEEEtran.bst: No hyphenation pattern has been}%
\typeout{** loaded for the language `#1'. Using the pattern for}%
\typeout{** the default language instead.}%
\else
\language=\csname l@#1\endcsname
\fi
#2}}
\providecommand{\BIBdecl}{\relax}
\BIBdecl

\bibitem{neshenko_demystifying_2019}
N.~Neshenko, E.~{Bou-Harb}, J.~Crichigno, G.~Kaddoum, and N.~Ghani,
  ``Demystifying {{IoT Security}}: {{An Exhaustive Survey}} on {{IoT
  Vulnerabilities}} and a {{First Empirical Look}} on {{Internet}}-{{Scale IoT
  Exploitations}},'' \emph{IEEE Communications Surveys Tutorials}, vol.~21,
  no.~3, pp. 2702--2733, thirdquarter 2019.

\bibitem{wang_physical-layer_2016}
X.~Wang, P.~Hao, and L.~Hanzo, ``Physical-layer authentication for wireless
  security enhancement: Current challenges and future developments,''
  \emph{IEEE Communications Magazine}, vol.~54, no.~6, pp. 152--158, Jun. 2016.

\bibitem{wang_wireless_2016}
W.~Wang, Z.~Sun, S.~Piao, B.~Zhu, and K.~Ren, ``Wireless {{Physical}}-{{Layer
  Identification}}: {{Modeling}} and {{Validation}},'' \emph{IEEE Transactions
  on Information Forensics and Security}, vol.~11, no.~9, pp. 2091--2106, Sep.
  2016.

\bibitem{yu_blind_2016}
J.~Yu, A.~Hu, and L.~Peng, ``Blind {{DCTF}}-based estimation of carrier
  frequency offset for {{RF}} fingerprint extraction,'' in \emph{2016 8th
  {{International Conference}} on {{Wireless Communications Signal Processing}}
  ({{WCSP}})}, Oct. 2016, pp. 1--6.

\bibitem{brik_wireless_2008}
V.~Brik, S.~Banerjee, M.~Gruteser, and S.~Oh, ``Wireless device identification
  with radiometric signatures,'' in \emph{Proceedings of the 14th {{ACM}}
  International Conference on {{Mobile}} Computing and Networking}.\hskip 1em
  plus 0.5em minus 0.4em\relax {ACM}, 2008, pp. 116--127.

\bibitem{danev_transient-based_2009}
B.~Danev and S.~Capkun, ``Transient-based identification of wireless sensor
  nodes,'' in \emph{2009 {{International Conference}} on {{Information
  Processing}} in {{Sensor Networks}}}, Apr. 2009, pp. 25--36.

\bibitem{vo-huu_fingerprinting_2016}
T.~D. {Vo-Huu}, T.~D. {Vo-Huu}, and G.~Noubir,
  ``\BIBforeignlanguage{en}{Fingerprinting {{Wi}}-{{Fi Devices Using Software
  Defined Radios}}},'' in \emph{\BIBforeignlanguage{en}{Proceedings of the 9th
  {{ACM Conference}} on {{Security}} \& {{Privacy}} in {{Wireless}} and
  {{Mobile Networks}} - {{WiSec}} '16}}.\hskip 1em plus 0.5em minus 0.4em\relax
  {Darmstadt, Germany}: {ACM Press}, 2016, pp. 3--14.

\bibitem{ren_practical_2018}
Y.~Ren, L.~Peng, W.~Bai, and J.~Yu, ``A {{Practical Study Of Channel Influence
  On Radio Frequency Fingerprint Features}},'' in \emph{2018 {{IEEE
  International Conference}} on {{Electronics}} and {{Communication
  Engineering}} ({{ICECE}})}, Dec. 2018, pp. 1--7.

\bibitem{peng_design_2019}
L.~Peng, A.~Hu, J.~Zhang, Y.~Jiang, J.~Yu, and Y.~Yan, ``Design of a {{Hybrid
  RF Fingerprint Extraction}} and {{Device Classification Scheme}},''
  \emph{IEEE Internet of Things Journal}, vol.~6, no.~1, pp. 349--360, Feb.
  2019.

\bibitem{chatterjee_rf-puf_2018}
B.~Chatterjee, D.~Das, and S.~Sen, ``\BIBforeignlanguage{en}{{{RF}}-{{PUF}}:
  {{IoT}} security enhancement through authentication of wireless nodes using
  in-situ machine learning},'' in \emph{\BIBforeignlanguage{en}{2018 {{IEEE
  International Symposium}} on {{Hardware Oriented Security}} and {{Trust}}
  ({{HOST}})}}.\hskip 1em plus 0.5em minus 0.4em\relax {Washington, DC}:
  {IEEE}, Apr. 2018, pp. 205--208.

\bibitem{zhou_design_2019}
X.~Zhou, A.~Hu, G.~Li, L.~Peng, Y.~Xing, and J.~Yu, ``Design of a {{Robust RF
  Fingerprint Generation}} and {{Classification Scheme}} for {{Practical Device
  Identification}},'' in \emph{2019 {{IEEE Conference}} on {{Communications}}
  and {{Network Security}} ({{CNS}})}, Jun. 2019, pp. 196--204.

\bibitem{xiao_using_2008}
L.~Xiao, L.~J. Greenstein, N.~B. Mandayam, and W.~Trappe, ``Using the physical
  layer for wireless authentication in time-variant channels,'' \emph{IEEE
  Transactions on Wireless Communications}, vol.~7, no.~7, pp. 2571--2579, Jul.
  2008.

\bibitem{senigagliesi_statistical_2019}
L.~Senigagliesi, M.~Baldi, and E.~Gambi, ``\BIBforeignlanguage{en}{Statistical
  and {{Machine Learning}}-based {{Decision Techniques}} for {{Physical Layer
  Authentication}}},'' \emph{\BIBforeignlanguage{en}{arXiv:1909.07969 [cs]}},
  Sep. 2019.

\bibitem{zhang_physical_2019}
P.~Zhang, T.~Taleb, X.~Jiang, and B.~Wu, ``Physical {{Layer Authentication}}
  for {{Massive MIMO Systems}} with {{Hardware Impairments}},'' \emph{IEEE
  Transactions on Wireless Communications}, pp. 1--1, 2019.

\bibitem{rehman_analysis_2012}
S.~U. Rehman, K.~Sowerby, and C.~Coghill, ``Analysis of receiver front end on
  the performance of {{RF}} fingerprinting,'' in \emph{2012 {{IEEE}} 23rd
  {{International Symposium}} on {{Personal}}, {{Indoor}} and {{Mobile Radio
  Communications}} - ({{PIMRC}})}, Sep. 2012, pp. 2494--2499.

\bibitem{riyaz_deep_2018}
S.~Riyaz, K.~Sankhe, S.~Ioannidis, and K.~Chowdhury, ``Deep {{Learning
  Convolutional Neural Networks}} for {{Radio Identification}},'' \emph{IEEE
  Communications Magazine}, vol.~56, no.~9, pp. 146--152, Sep. 2018.

\bibitem{yu_robust_2019}
J.~Yu, A.~Hu, G.~Li, and L.~Peng, ``A {{Robust RF Fingerprinting Approach Using
  Multi}}-{{Sampling Convolutional Neural Network}},'' \emph{IEEE Internet of
  Things Journal}, pp. 1--1, 2019.

\bibitem{gopalakrishnan_robust_2019}
S.~Gopalakrishnan, M.~Cekic, and U.~Madhow, ``\BIBforeignlanguage{en}{Robust
  {{Wireless Fingerprinting}} via {{Complex}}-{{Valued Neural Networks}}},''
  \emph{\BIBforeignlanguage{en}{arXiv:1905.09388 [cs, eess, stat]}}, May 2019.

\bibitem{baldini_comparison_2019}
G.~Baldini, C.~Gentile, R.~Giuliani, and G.~Steri, ``Comparison of techniques
  for radiometric identification based on deep convolutional neural networks,''
  \emph{Electronics Letters}, vol.~55, no.~2, pp. 90--92, 2019.

\bibitem{agadakos_deep_2019}
I.~Agadakos, N.~Agadakos, J.~Polakis, and M.~R. Amer,
  ``\BIBforeignlanguage{en}{Deep {{Complex Networks}} for
  {{Protocol}}-{{Agnostic Radio Frequency Device Fingerprinting}} in the
  {{Wild}}},'' \emph{\BIBforeignlanguage{en}{arXiv:1909.08703 [cs, eess]}},
  Sep. 2019.

\bibitem{wu_deep_2018}
Q.~Wu, C.~Feres, D.~Kuzmenko, D.~Zhi, Z.~Yu, X.~Liu, and X.~`Leo'~Liu,
  ``\BIBforeignlanguage{en}{Deep learning based {{RF}} fingerprinting for
  device identification and wireless security},''
  \emph{\BIBforeignlanguage{en}{Electronics Letters}}, vol.~54, no.~24, pp.
  1405--1407, Nov. 2018.

\bibitem{merchant_deep_2018-1}
K.~Merchant, S.~Revay, G.~Stantchev, and B.~Nousain, ``Deep {{Learning}} for
  {{RF Device Fingerprinting}} in {{Cognitive Communication Networks}},''
  \emph{IEEE Journal of Selected Topics in Signal Processing}, vol.~12, no.~1,
  pp. 160--167, Feb. 2018.

\bibitem{hanna_icnc_2019}
S.~S. Hanna and D.~Cabric, ``Deep {{Learning Based Transmitter Identification}}
  using {{Power Amplifier Nonlinearity}},'' in \emph{2019 {{International
  Conference}} on {{Computing}}, {{Networking}} and {{Communications}}
  ({{ICNC}})}, Feb. 2019, pp. 674--680.

\bibitem{youssef_machine_2017}
K.~Youssef, L.-S. Bouchard, K.~Z. Haigh, H.~Krovi, J.~Silovsky, and C.~P.~V.
  Valk, ``Machine {{Learning Approach}} to {{RF Transmitter Identification}},''
  \emph{arXiv:1711.01559 [cs, eess, stat]}, Nov. 2017.

\bibitem{gritsenko_finding_2019}
A.~Gritsenko, Z.~Wang, T.~Jian, J.~Dy, K.~Chowdhury, and S.~Ioannidis,
  ``\BIBforeignlanguage{en}{Finding a `{{New}}' {{Needle}} in the {{Haystack}}:
  {{Unseen Radio Detection}} in {{Large Populations Using Deep Learning}}},''
  in \emph{\BIBforeignlanguage{en}{2019 {{IEEE International Symposium}} on
  {{Dynamic Spectrum Access Networks}} ({{DySPAN}})}}.\hskip 1em plus 0.5em
  minus 0.4em\relax {Newark, NJ, USA}: {IEEE}, Nov. 2019, pp. 1--10.

\bibitem{hanna_spawc_2020}
S.~Hanna, S.~Karunaratne, and D.~Cabric, ``Deep {{Learning Approaches}} for
  {{Open Set Wireless Transmitter Authorization}},'' in \emph{2020 {{IEEE}}
  21st International Workshop on Signal Processing Advances in Wireless
  Communications ({{SPAWC}}) ({{IEEE SPAWC}} 2020)}, {Atlanta, USA}, May 2020,
  accepted for publication.

\bibitem{openset_survey_2019}
C.~Geng, S.-j. Huang, and S.~Chen, ``\BIBforeignlanguage{en}{Recent
  {{Advances}} in {{Open Set Recognition}}: {{A Survey}}},''
  \emph{\BIBforeignlanguage{en}{arXiv:1811.08581 [cs, stat]}}, Jul. 2019.

\bibitem{anomaly_detection_survey_2019}
R.~Chalapathy and S.~Chawla, ``\BIBforeignlanguage{en}{Deep {{Learning}} for
  {{Anomaly Detection}}: {{A Survey}}},''
  \emph{\BIBforeignlanguage{en}{arXiv:1901.03407 [cs, stat]}}, Jan. 2019.

\bibitem{xie_blind_2018}
N.~Xie and S.~Zhang, ``Blind {{Authentication}} at the {{Physical Layer Under
  Time}}-{{Varying Fading Channels}},'' \emph{IEEE Journal on Selected Areas in
  Communications}, vol.~36, no.~7, pp. 1465--1479, Jul. 2018.

\bibitem{gu_physical_2020}
Z.~Gu, H.~Chen, P.~Xu, Y.~Li, and B.~Vucetic,
  ``\BIBforeignlanguage{en}{Physical {{Layer Authentication}} for
  {{Non}}-coherent {{Massive SIMO}}-{{Based Industrial IoT Communications}}},''
  \emph{\BIBforeignlanguage{en}{arXiv:2001.07315 [eess]}}, Jan. 2020.

\bibitem{xu_device_2016-1}
Q.~Xu, R.~Zheng, W.~Saad, and Z.~Han, ``Device {{Fingerprinting}} in {{Wireless
  Networks}}: {{Challenges}} and {{Opportunities}},'' \emph{IEEE Communications
  Surveys Tutorials}, vol.~18, no.~1, pp. 94--104, Firstquarter 2016.

\bibitem{fang_learning-aided_2018}
H.~Fang, X.~Wang, and L.~Hanzo, ``\BIBforeignlanguage{en}{Learning-{{Aided
  Physical Layer Authentication}} as an {{Intelligent Process}}},''
  \emph{\BIBforeignlanguage{en}{arXiv:1808.02456 [cs]}}, Aug. 2018.

\bibitem{nguyen_device_2011}
N.~T. Nguyen, G.~Zheng, Z.~Han, and R.~Zheng, ``Device fingerprinting to
  enhance wireless security using nonparametric {{Bayesian}} method,'' in
  \emph{2011 {{Proceedings IEEE INFOCOM}}}, Apr. 2011, pp. 1404--1412.

\bibitem{gulati_gmm_2013}
N.~Gulati, R.~Greenstadt, K.~R. Dandekar, and J.~M. Walsh, ``{{GMM Based
  Semi}}-{{Supervised Learning}} for {{Channel}}-{{Based Authentication
  Scheme}},'' in \emph{2013 {{IEEE}} 78th {{Vehicular Technology Conference}}
  ({{VTC Fall}})}, Sep. 2013, pp. 1--6.

\bibitem{weinand_application_2017}
A.~Weinand, M.~Karrenbauer, R.~Sattiraju, and H.~D. Schotten,
  ``\BIBforeignlanguage{en}{Application of {{Machine Learning}} for {{Channel}}
  based {{Message Authentication}} in {{Mission Critical Machine Type
  Communication}}},'' \emph{\BIBforeignlanguage{en}{arXiv:1711.05088 [cs,
  eess]}}, Nov. 2017.

\bibitem{sankhe_oracle_2018}
K.~Sankhe, M.~Belgiovine, F.~Zhou, S.~Riyaz, S.~Ioannidis, and K.~Chowdhury,
  ``{{ORACLE}}: {{Optimized Radio clAssification}} through {{Convolutional
  neuraL nEtworks}},'' \emph{arXiv:1812.01124 [cs, eess]}, Dec. 2018.

\bibitem{sankhe_no_2019}
K.~Sankhe, M.~Belgiovine, F.~Zhou, L.~Angioloni, F.~Restuccia, S.~D'Oro,
  T.~Melodia, S.~Ioannidis, and K.~Chowdhury, ``No {{Radio Left Behind}}:
  {{Radio Fingerprinting Through Deep Learning}} of {{Physical}}-{{Layer
  Hardware Impairments}},'' \emph{IEEE Transactions on Cognitive Communications
  and Networking}, pp. 1--1, 2019.

\bibitem{restuccia_deepradioid_2019}
F.~Restuccia, S.~D'Oro, A.~{Al-Shawabka}, M.~Belgiovine, L.~Angioloni,
  S.~Ioannidis, K.~Chowdhury, and T.~Melodia, ``{{DeepRadioID}}:
  {{Real}}-{{Time Channel}}-{{Resilient Optimization}} of {{Deep
  Learning}}-based {{Radio Fingerprinting Algorithms}},''
  \emph{arXiv:1904.07623 [cs]}, Apr. 2019.

\bibitem{baldini_physical_2019}
G.~Baldini, R.~Giuliani, and F.~Dimc, ``\BIBforeignlanguage{en}{Physical layer
  authentication of {{Internet}} of {{Things}} wireless devices using
  convolutional neural networks and recurrence plots},''
  \emph{\BIBforeignlanguage{en}{Internet Technology Letters}}, vol.~2, no.~2,
  p. e81, 2019.

\bibitem{pan_specific_2019}
Y.~Pan, S.~Yang, H.~Peng, T.~Li, and W.~Wang, ``Specific {{Emitter
  Identification Based}} on {{Deep Residual Networks}},'' \emph{IEEE Access},
  vol.~7, pp. 54\,425--54\,434, 2019.

\bibitem{ding_specific_2018-1}
L.~Ding, S.~Wang, F.~Wang, and W.~Zhang, ``Specific {{Emitter Identification}}
  via {{Convolutional Neural Networks}},'' \emph{IEEE Communications Letters},
  vol.~22, no.~12, pp. 2591--2594, Dec. 2018.

\bibitem{jafari_iot_2018}
H.~Jafari, O.~Omotere, D.~Adesina, H.~Wu, and L.~Qian, ``{{IoT Devices
  Fingerprinting Using Deep Learning}},'' in \emph{{{MILCOM}} 2018 - 2018
  {{IEEE Military Communications Conference}} ({{MILCOM}})}, Oct. 2018, pp.
  1--9.

\bibitem{morin_transmitter_2019}
C.~Morin, L.~Cardoso, J.~Hoydis, J.-M. Gorce, and T.~Vial,
  ``\BIBforeignlanguage{en}{Transmitter {{Classification With Supervised Deep
  Learning}}},'' \emph{\BIBforeignlanguage{en}{arXiv:1905.07923 [cs, eess]}},
  May 2019.

\bibitem{yu_radio_2019}
J.~Yu, A.~Hu, F.~Zhou, Y.~Xing, Y.~Yu, G.~Li, and L.~Peng, ``Radio {{Frequency
  Fingerprint Identification Based}} on {{Denoising Autoencoders}},''
  \emph{arXiv:1907.08809 [cs, eess]}, Jul. 2019.

\bibitem{wong_emitter_2018}
L.~J. Wong, W.~C. Headley, and A.~J. Michaels,
  ``\BIBforeignlanguage{en}{Emitter {{Identification Using CNN IQ Imbalance
  Estimators}}},'' \emph{\BIBforeignlanguage{en}{arXiv:1808.02369 [eess]}},
  Aug. 2018.

\bibitem{orbit_2005}
D.~Raychaudhuri, I.~Seskar, M.~Ott, S.~Ganu, K.~Ramachandran, H.~Kremo,
  R.~Siracusa, H.~Liu, and M.~Singh, ``Overview of the {{ORBIT}} radio grid
  testbed for evaluation of next-generation wireless network protocols,'' in
  \emph{Wireless {{Communications}} and {{Networking Conference}}, 2005
  {{IEEE}}}, vol.~3.\hskip 1em plus 0.5em minus 0.4em\relax {IEEE}, 2005, pp.
  1664--1669.

\bibitem{shu_doc_2017}
L.~Shu, H.~Xu, and B.~Liu, ``\BIBforeignlanguage{en}{{{DOC}}: {{Deep Open
  Classification}} of {{Text Documents}}},''
  \emph{\BIBforeignlanguage{en}{arXiv:1709.08716 [cs]}}, Sep. 2017.

\bibitem{openmax_2015}
A.~Bendale and T.~Boult, ``\BIBforeignlanguage{en}{Towards {{Open Set Deep
  Networks}}},'' \emph{\BIBforeignlanguage{en}{arXiv:1511.06233 [cs]}}, Nov.
  2015.

\bibitem{kotz_extreme_2000}
S.~Kotz and S.~Nadarajah, \emph{\BIBforeignlanguage{en}{Extreme {{Value
  Distributions}}: {{Theory}} and {{Applications}}}}.\hskip 1em plus 0.5em
  minus 0.4em\relax {World Scientific}, 2000.

\bibitem{yoshihashi_classification-reconstruction_2019}
R.~Yoshihashi, W.~Shao, R.~Kawakami, S.~You, M.~Iida, and T.~Naemura,
  ``Classification-{{Reconstruction Learning}} for {{Open}}-{{Set
  Recognition}},'' in \emph{2019 {{IEEE}}/{{CVF Conference}} on {{Computer
  Vision}} and {{Pattern Recognition}} ({{CVPR}})}, Jun. 2019, pp. 4011--4020.

\bibitem{cekic_robust_2020}
M.~Cekic, S.~Gopalakrishnan, and U.~Madhow, ``\BIBforeignlanguage{en}{Robust
  {{Wireless Fingerprinting}}: {{Generalizing Across Space}} and {{Time}}},''
  \emph{\BIBforeignlanguage{en}{arXiv:2002.10791 [cs, eess, stat]}}, Feb. 2020.

\bibitem{al-shawabka_exposing_2020}
A.~{Al-Shawabka}, F.~Restuccia, S.~D'Oro, T.~Jian, B.~C. Rendon, N.~Soltani,
  J.~Dy, K.~Chowdhury, S.~Ioannidis, and T.~Melodia,
  ``\BIBforeignlanguage{en}{Exposing the {{Fingerprint}}: {{Dissecting}} the
  {{Impact}} of the {{Wireless Channel}} on {{Radio Fingerprinting}}},''
  \emph{\BIBforeignlanguage{en}{Proc. of IEEE Conference on Computer
  Communications (INFOCOM)}}, p.~10, 2020.

\bibitem{he_deep_2015}
K.~He, X.~Zhang, S.~Ren, and J.~Sun, ``Deep {{Residual Learning}} for {{Image
  Recognition}},'' \emph{arXiv:1512.03385 [cs]}, Dec. 2015.

\bibitem{sun_classification_2009}
Y.~Sun, A.~K.~C. Wong, and M.~S. Kamel,
  ``\BIBforeignlanguage{en}{{{CLASSIFICATION OF IMBALANCED DATA}}: {{A
  REVIEW}}},'' \emph{\BIBforeignlanguage{en}{International Journal of Pattern
  Recognition and Artificial Intelligence}}, vol.~23, no.~04, pp. 687--719,
  Jun. 2009.

\end{thebibliography}

\begin{IEEEbiographynophoto}{Samer Hanna}
received the B.Sc. degree in Electrical Engineering from Alexandria University, Alexandria, Egypt in 2013, and the M.Sc. degree in Engineering Mathematics from the same university in 2017. He is currently  pursuing  the  Ph.D.  at  the  University  of  California,  Los Angeles, CA, USA. His research interests includes the applications of machine learning  in wireless communications and coordinated communications using  unmanned aerial vehicles .
\end{IEEEbiographynophoto}

\begin{IEEEbiographynophoto}{Samurdhi Karunaratne}
received the B.Sc. degree in Computer Engineering from the University of Peradeniya, Sri Lanka, in 2019. Currently he is an M.S./Ph.D. student at the University of California, Los Angeles working in the Cognitive Reconfigurable Embedded Systems Lab (CORES). His research interests include optimization algorithms and the applications of machine learning techniques in wireless communications.
\end{IEEEbiographynophoto}

\begin{IEEEbiographynophoto}{Danijela Cabric}
Danijela Cabric is Professor in Electrical  and  Computer  Engineering  at  University  of  California,  Los  Angeles.  She  earned  MS  degree  in  Electrical  Engineering in 2001, UCLA and Ph.D. in Elec-trical Engineering in 2007, UC Berkeley, Dr. Cabric received the Samueli Fellowship in 2008, the Okawa Foundation Research Grant in 2009, Hellman Fel-lowship in 2012 and the National Science Foundation Fac-ulty Early Career Development (CAREER) Award in 2012. She  is  now  Associated  Editor  of  IEEE  Transactions  of  Cognitive Communications and Networking and IEEE Transactions of Wireless Communications. Her research interests  include  novel  radio  architecture,  signal  pro-cessing, and networking techniques for cognitive radio, 5G and massive MIMO systems. She is Senior Member of IEEE and IEEE ComSoc Distinguished Lecturer.
\end{IEEEbiographynophoto}

\end{document}